\documentstyle[prl,aps,preprint,psfig]{revtex}

\begin{document}
\draft 

\title{Hopf invariant for long-wavelength skyrmions 
       in quantum Hall systems for integer and fractional fillings}

\author{K. Sengupta and Victor M. Yakovenko}

\address{Department of Physics and Center for Superconductivity,
	University of Maryland, College Park, MD 20742-4111}

\date{\today }

\maketitle

\begin{abstract}

We show that a Hopf term exists in the effective action of
long-wavelength skyrmions in quantum Hall systems for both odd integer
and fractional filling factors $\nu = 1/(2s +1)$, where $s$ is an
integer. We evaluate the prefactor of the Hopf term using Green
function method in the limit of strong external magnetic field using
model of local interaction. The prefactor ($N$) of the Hopf term is
found to be equal to $\nu$. The spin and charge densities and hence
the total spin and charge of the skyrmion are computed from the
effective action.  The total spin is found to have a dominant
contribution from the Berry term in the effective action and to
increase with the size of the skyrmion. The charge and the statistics
of the skyrmion, on the other hand, are completely determined by the
prefactor of the Hopf term.  Consequently, the skyrmions have charge
$\nu e$ and are Fermions (anyons) for odd integer (fractional)
fillings. We also obtain the effective action of the skyrmions at
finite temperature. It is shown that at finite temperature, the value
of the prefactor of the Hopf term depends on the order in which the
zero-momentum and zero-frequency limits are taken.

\end{abstract}


\section {Introduction}
\label{secint}

Two-dimensional (2D) electron gases are known to exhibit a rich
variety of interesting phenomena with variation of particle density or
magnetic field. The most striking among these phenomena are the
integer and fractional quantum Hall effects
\cite{Prange,Stone,Chakraborty,Jan,DasSarma}. Since these phenomena
take place at very high magnetic field, it may seem reasonable to
think that the presence of a Zeeman term in the Hamiltonian would
preclude any dynamics for the spin degree of freedom. This would have
indeed been the case if the Lande g factor for the electrons in the
sample were close to the free electron Lande g factor. The
lowest-lying excitations of the system would then be
quasihole-quasiparticle pairs with opposite spins. However, as
observed by Halperin \cite{Halperin}, the Lande g factor for electrons
in Ga-As samples is much lower than the free-electron Lande g
factor. The Zeeman splitting for the electrons in these systems is,
therefore, small compared to both the cyclotron energy and the typical
Coulomb energy. As a result, the spin degrees of freedom become
dynamical, in spite of the presence of a strong external magnetic
field. Nevertheless, for certain filling fractions ($\nu = 1,1/3$),
the ground state of the system is ferromagnetic even in the limit of
vanishing Zeeman energy because of exchange interaction between the
electrons. For these filling fractions, the lowest lying excitations
are topologically non-trivial spin configurations called skyrmions
\cite{DasSarma,Sondhi,Fertig}. The spatial extent of such
configurations or skyrmions is determined by the relative strength of
the Coulomb and Zeeman energies in the system. When the Zeeman energy
is negligible compared to the Coulomb energy, the skyrmions have a
large radius and are referred to as long-wavelength skyrmions. A small
deviation from a ferromagnetic filling fraction creates such
skyrmions with many electron spins reversed which strongly reduces the
spin polarization of the system.  A clear signature of such spin
depolarization, suggesting skyrmionic spin configurations, has been
observed experimentally at filling factor $\nu = 1$
\cite{Kukushkin,Barret}. The experimental realization of skyrmions for
fractional filling factors is much more difficult since it requires a
very low Lande g factor. Nevertheless, recently Leadly ${\it et\,al.}$
\cite{Leadly} have observed signature of skyrmions at $\nu = 1/3$ by
reducing the Lande g factor of the sample by applying external
pressure.

The spin configurations of the system can be characterized by a unit
vector field ${\bf d}(x,y,t)$ that gives the direction of the local
spin. For long-wavelength skyrmions, the space-time modulation of the
spin configuration and hence the ${\bf d}$ field is slow. Thus it is
possible to derive a low-energy effective action for the skyrmions in
terms of ${\bf d}$ field gradients. It is well known
\cite{Sondhi,Fertig,Moon} that in quantum Hall ferromagnets,
such a low energy effective action $S_{\rm FM}$ would contain the
Berry and the usual gradient terms:
                \begin{mathletters}
\begin{equation}       
S_{\rm FM} = S_{\rm Berry} + S_{\rm E}, 
\label{ferroa}
\end{equation}
\begin{equation}    
S_{\rm Berry} = \frac{\rho_0}{2} 
\int d^2r\,dt\, B_0,
\label{ferrob}
\end{equation}    
\begin{equation}    
S_{\rm E} = -\frac{\kappa}{2} \int d^2r\,dt 
\left( {\nabla} {\bf d} \right)^2 ,
\label{ferroc}
\end{equation}
                    \end{mathletters}
where $B_0$ is defined in Eq.\ (\ref{field tensor}), $\rho_0$ is the
ground state density and  
$\kappa$ is the spin stiffness constant. This is to be contrasted with
the effective action for antiferromagnetic systems, 
which contains a term quadratic in time derivative of the unit vector
${\bf d}$ as shown by Wen and Zee \cite{Zee} and Haldane
\cite{Haldane}.

If the ground state of the system does not have time-reversal and
parity invariance, the effective action may also contain the Hopf term
given by
             \begin{eqnarray}
S_{\rm Hopf}&=& \frac{N}{32 \pi} \int d^2 {\bf r}\,dt \,
\epsilon^{\mu \nu \lambda} B_{\mu} F_{\nu \lambda},
\label{Hopfterm} 
              \end {eqnarray}
where $\epsilon^{\mu \nu \lambda}$ is the completely antisymmetric
tensor, the prefactor $N$ is a topological invariant and 
$B_{\mu}$ is an auxiliary gauge field given by
                 \begin{eqnarray}
F_{\mu \nu} &=& \partial_{\mu}B_{\nu} 
- \partial_{\nu} B_{\mu} 
= {\bf d}\cdot (\partial_{\mu}{\bf d} 
\times \partial_{\nu}{\bf d} ). 
\label{field tensor}
                 \end{eqnarray}
For antiferromagnetic systems,
Wilczek and Zee have shown that the topological invariant $N$
determines the spin and the statistics of the skyrmion
\cite{Wilczek}. If $N$ is an odd (even) integer, the skyrmion is a
Fermion (boson) with spin $N/2$. For fractional $N$, the skyrmion is
an anyon. However, as we shall see in Sec.\ \ref{secstat}, the method
of Ref.\ \cite{Wilczek} leads to quite different conclusions for the
spin of the skyrmions in ferromagnetic systems due to the presence of
the Berry term (\ref{ferrob}) in the effective action. The charge and the
statistics, on the other hand, are still completely determined by the
topological invariant $N$.

The presence of a Hopf term in the effective action of systems
described by a unit vector field ${\bf d}$ is well known in many
different areas of condensed matter. It was initially conjectured by
Dzyaloshinskii ${\it et\,al.}$ that this term may exist in the
effective action of planar antiferromagnets \cite{Dzyaloshinskii}.
However, microscopic calculations showed that such a term is absent in
the effective action of these systems because of symmetry requirements
\cite{Zee,Haldane}. There are, nonetheless, various other systems
where symmetry requirements permit a non-zero topological term in the
low-energy effective action. A few examples are ${\rm He}^3$-A films
\cite{Volovik1}, lattice quantum Hall systems \cite{Yakovenko1} and
quasi-one-dimensional organic conductors \cite{Yakovenko2}. Recently,
it has been suggested that the superconducting phase in $\rm
Sr_2RuO_4$ has a spin-triplet $p$-wave pairing state with broken
parity and time-reversal symmetry similar to ${\rm He}^3$-A films
\cite{Maeno}. Consequently, the corresponding low-energy effective
action of $\rm Sr_2RuO_4$ is also expected to have similar topological
term \cite{Sigrist}.

Volovik and Yakovenko \cite{Volovik1,Yakovenko1} have
derived an explicit expression for the 
topological invariant $N$ for a general class of mean-field Fermionic
models, with the electron Hamiltonian of the form
               \begin{equation}
{\hat H} = {\hat H}_0 + {\bbox \sigma} \cdot {\bf d} ({\bf r},t) 
{\hat H}_1,
\label{Hamiltonian}
                \end{equation}
where ${\bbox \sigma}$ are the Pauli matrices acting on the spin
indices of the electrons. This class includes the above mentioned
systems \cite{Volovik1,Yakovenko1,Yakovenko2,Maeno}. It was shown that
the expression for the topological invariant $N$ is
                 \begin{eqnarray}
N &=& 2 \pi {\rm Tr }
\left(\frac{\partial G_0^{-1}}{\partial \omega} 
G_0 \frac{\partial G_0^{-1}}{\partial k_x} G_0
\frac{\partial G_0^{-1}}{\partial k_y} G_0 \right),
\label{invar}
                 \end{eqnarray}
where Tr denotes all integrations and matrix traces, $k_x, k_y$ and
$\omega$ are the electron momenta and frequency, and 
$G_0(k_x,k_y,\omega) = (\omega - {\hat H}_0 -\sigma_z {\hat H}_1 +
i\eta {\rm Sgn} \omega)^{-1}$ is the unperturbed electron Green
function. In principle, this treatment should also hold for quantum Hall
systems, since their Hamiltonians can be cast into the same class of
mean-field models (\ref{Hamiltonian}).
However, since $k_x$ and $k_y$ are no longer good quantum numbers
simultaneously in the presence of magnetic field, Eq.\ (\ref{invar}) cannot
be directly applied to the present case.  Nevertheless, if we compare
the expression of the topological invariant $N$ to that of Hall
conductivity in quantum Hall systems \cite{Niu},
               \begin{eqnarray}
\frac{h}{e^2}\sigma_{xy} &=& 2 \pi {\rm Tr }
\left(\frac{\partial G_0^{-1}}{\partial \omega} 
G_0 \frac{\partial G_0^{-1}}{\partial \phi_x} G_0
\frac{\partial G_0^{-1}}{\partial \phi_y} G_0 \right),
\label{Hall}
                \end{eqnarray}
we find that expressions (\ref{invar}) and (\ref{Hall}) are similar
except that the momenta ($k_x,k_y$) are replaced by boundary phase
parameters ($\phi_x,\phi_y$) and the integration over momenta are to
be replaced by average over the boundary phases \cite{comment}. In
fact, as we shall 
see, $N$ is the same topological invariant that determines the Hall
conductivity in quantum Hall systems, and properly modified, the
formalism developed in Ref. \cite{Volovik1} gives exactly the same
Eq.\ (\ref{Hall}) for $N$.

Recently, there has been a lot of interest in finding the Hopf term in
the effective action of long-wavelength skyrmions in quantum Hall
systems at $\nu = 1$. Some of these works \cite{Apel,Ray} use
lowest Landau level (LLL) projection technique to derive the Hopf
term. However, it has been pointed out \cite{Volovik2} that the
parameterization of the unit vector ${\bf d}$ in terms of the Euler
angles in Ref.\cite{Apel} is questionable. This gave rise to a subsequent
controversy \cite{Apel2,Volovik3}, which we aim to resolve.  Further, 
the validity of the LLL projection in the present context was also
questioned by Iordanskii and Plyasunov
\cite{Iordanskii,Iord}, who derived the prefactor of the Hopf term 
by explicit term by term evaluation of the effective action,
starting from a mean-field Hamiltonian but without any LLL
projection. A similar work, avoiding the LLL projection was later done
by Ray \cite{Ray2}. Although the end result for the prefactor $N$ is
the same in all these works \cite{Apel,Ray,Iordanskii,Iord,Ray2}, it
is not quite clear whether the result is an artifact of the simplicity
of the models or assumptions used. The aim of the present work is to
point out the robustness of this result and to generalize it for
fractional filling factors $\nu = 1/(2s+1)$, where s is an
integer. We also explicitly compute the spin and charge densities of
the skyrmion, and hence its total spin and charge. It is shown that the
dominant contribution to the skyrmion spin comes from the Berry term
in the action. The contribution to the skyrmion spin from the Hopf
term comes equally from the bulk and the edge, but is small compared
to the contribution of the Berry term.  This result coincides with the
results obtained by Baez ${\it et\,al.}$ on a phenomenological basis
\cite{Baez}. The statistics of the skyrmion is also computed using the
method of Ref. \cite{Wilczek} and is found to be determined by the
prefactor of the Hopf term. Our result regarding the statistics
corroborates the result obtained by Yang and Sondhi \cite{Yang}
using a variational wave-function for the skyrmion, but differs from
that of Dziarmaga \cite{Dziarmaga}.

The fate of the Hopf term at finite temperature is another very
interesting question. At zero temperature, it is
possible to carry out a derivative expansion of the polarization
tensor and thus obtain an expression for the prefactor of the Hopf
term in terms of the Green functions. However the situation is much
more tricky at finite temperature. It is well known
\cite{Das,Aitchison,Kao,Goan}, at least in the case of U(1) gauge
fields (electromagnetic field), that the polarization tensor becomes a
non-analytic function of frequency and momentum at finite temperature
and a derivative expansion can not be carried out unambiguously. The
zero-frequency zero-momentum limit ($ p_0 \rightarrow 0, {\bf
p}\rightarrow 0$) depends on the order in which the limits are
taken. This order is generally chosen from physical consideration and
depends on the system of interest. In particular, in quantum Hall
systems, the prefactor of the Chern-Simons term (the
equivalent of the Hopf term for Abelian gauge fields) which determines
the Hall conductivity depends on the order in which the zero-frequency
and the zero-momentum limits are taken. In this work, we use imaginary
time Matsubara formalism to derive an effective action for
skyrmions at finite temperature and $\nu = 1$. We show that, analogous
to the Chern-Simons term for the Abelian gauge fields, it is in
general not possible to obtain a local Hopf term in position space at
finite temperature. Nevertheless, one can still obtain a rather simple
effective action for skyrmions in momentum space. We evaluate the
prefactor of the Hopf term in both the dynamic (${\bf p}/p_o
\to 0$) and the static ($p_o/{\bf p} \to  0$)
limit. In the static limit, the prefactor depends on temperature, 
while in the dynamic limit, it is independent of
temperature and has the same value as at zero temperature.

The organization of the paper is as follows. In Sec.\ref{secexp}, we
derive the effective action at zero temperature and obtain an
expression for the prefactor $N$ in terms of the Green functions. In
Sec.\ref{seceva}, the value of this prefactor is evaluated. In
Sec.\ref{secspi}, we derive expressions for spin and charge densities
and obtain the total spin and charge of the skyrmion. The statistics
of the skyrmion is obtained in Sec.\ref{secstat}. These results
are generalized for fractional fillings in
Sec.\ref{secfra}. In Sec.\ref{secfin}, we derive the effective
action at finite temperature and obtain the prefactor of the Hopf
term for both the static and the dynamic limits. This is followed by
conclusion in Sec.\ref{seccon}. Some details of the calculations are
sketched in Appendices A-C while in Appendix D, we present a list of
notations used, for clarity.

\section{Expression for $N$ at $\nu$ = 1}
\label{secexp}

In this section, we derive the effective action at zero temperature
and from there obtain an expression for the prefactor $N$ in terms
of the Green functions of the system at $\nu$ =1. Throughout the rest
of the work, natural units $\hbar = c = 1$ are used. The
convention of subscripts and superscripts are as follows. The Greek
letters are used to denote space-time indices and these can take
values (0,1,2) where 0 denotes time component and 1 and 2 denote the
two space directions. The letters $i$, $j$, and $k$ are used for the
indices of the Pauli matrices $\sigma_i$ and take values (1,2,3). 
The letters $a$, $b$, and $c$ denote
either space indices 1 and 2 or spinor indices $\uparrow$ and
$\downarrow$. All repeated indices are summed over unless
explicitly stated otherwise. We also use the following convention for
vectors and operators whenever needed. All contravariant vectors are
taken as $X^{\mu}$ = ($X^0,{\bf X}$) and all operators as
$\partial^{\mu}$ = ($\partial^0, -{\bbox \bigtriangledown}$). The
covariant counterparts of the contravariant vectors and operators are
obtained by applying the metric tensor $g_{\mu \nu} = g^{\mu \nu} $ =
diag(1,-1,-1) \cite{Jackson}.

The action for the system with a model local interaction $ V({\bf
r}_1-{\bf r}_2) = \gamma_0 \delta ({\bf r}_1-{\bf r}_2)$
\cite{comment1} can be written using Hartree-Fock approximation as
\cite{Iordanskii,Iord}

               \begin{equation}
S[\psi^{\dagger},\psi] = \int d^2{\bf r}\,dt\, 
\psi^{\dagger}_a ({\bf r},t) 
\left( i \partial_0  I - H + \epsilon_F I\right )_{ab} 
\psi_b ({\bf r},t),
\label{mean field}
               \end{equation}
where the Fermionic field $\psi$ is a two component spinor, I is the 2
$\times$ 2 unit matrix, $\epsilon_F$ is the Fermi energy and $H$ is
the mean-field Hamiltonian density of the system given by
               \begin{equation}
H = \frac {\left[ {\hat{\bf p}} -
 e {\bf A} ({\bf r}) \right ]^2} {2 m} I
 - \gamma_0 {\bf d} ({\bf r},t) \cdot {\bbox \sigma}.
\label{mfh}
                \end{equation}
Here ${\bf A}$ is the vector potential corresponding to the external
magnetic field ${\bf H}_0$, ${\bf d}$ is the unit vector field that
gives the direction of local spin, ${\bbox \sigma}$ are the Pauli
matrices, $\gamma_0 \sim e^2/l_B$ is the typical Coulomb energy of the
system, and $l_B = \sqrt{1/eH_0 }$ is the magnetic length. In this
treatment we shall neglect Zeeman energy ($E_z$) and also assume that
the characteristic cyclotron energy is much greater than all other
energy scales in the problem, ${\it i.e.}$ we consider the regime $
\omega_c\gg \gamma_0 \gg E_z $. This is the relevant regime for
long-wavelength skyrmions \cite{Apel,Iordanskii}.

To calculate the topological term in the action, it is convenient to
introduce 2 $\times$ 2 local unitary SU(2) rotation matrix $U({\bf
r},t)$, that corresponds to the local rotation of ${\bf d}({\bf
r},t)$ from the homogeneous field ${\bf d} = {\bf e_z} $
                  \begin{equation}
U ({\bf r},t)\sigma_z U^{-1}({\bf r},t) = 
{\bbox \sigma} \cdot {\bf d} ({\bf r},t).
                 \end{equation}
After the unitary transformation of the Fermi fields $\chi({\bf
 r},t) = U^{-1} ({\bf r},t) \psi ({\bf r},t) $, the action becomes
                  \begin{eqnarray}
S[\chi^{\dagger},\chi, Q_{\mu}^{\rm int}] 
&=& \int d^2 r\,dt\,\chi^{\dagger}_a ({\bf r},t)
\Bigg (  i \partial_0 I - Q_0^{\rm int} ({\bf r},t) \nonumber\\
&& - \frac{\left\{ \left[ {\hat{\bf p}} -  e {\bf A} ({\bf r}) 
\right] I - {{\bf Q}^{\rm int}} ({\bf r},t) \right \}^2 }{2 m} 
  + \gamma_0 \sigma_z + \epsilon_F I \Bigg )_{ab} 
\chi_b ({\bf r},t).
                 \end{eqnarray}
\noindent
The new spinor fields $\chi({\bf r},t)$ have their spin quantization
axis along ${\bf e_z}$, and $Q_{\mu}^{\rm int}$ ($\mu = 0,1,2$) are
the SU(2) gauge fields given by
                    \begin{eqnarray}
Q_{\mu}^{\rm int} &=& - i U^{-1} (\partial_{\mu} U)
 = \frac{1}{2} {\bbox \sigma} 
\cdot {\bbox \Omega}_{\mu}^{\rm int} = \frac{1}{2} \sigma_i 
\Omega_{\mu}^{i\,\rm int}.
                   \end{eqnarray}
The fields ${\bbox \Omega}_{\mu}^{\rm int}$ are pure gauge fields
satisfying the relation
                  \begin{eqnarray}
f_{\mu \nu}^{\rm int} &=&\partial_{\mu} 
{\bbox \Omega}_{\nu}^{\rm int} 
- \partial_{\nu} {\bbox \Omega}_{\mu}^{\rm int} 
- {\bbox \Omega}_{\mu}^{\rm int} 
\times {\bbox \Omega}_{\nu}^{\rm int} = 0.
\label{gauge}
                   \end{eqnarray}
Further, the rotation matrix $U$ satisfies the relation \cite{Volovik1}
                   \begin{equation}
i(\partial_{\mu} U) U^{-1} = \frac{1}{2} 
{\bbox \sigma} \cdot \left(- B_{\mu} {\bf d} 
+ {\bf d} \times \partial_{\mu} {\bf d}\right),
\label{rotation}
                  \end{equation}
where $B_{\mu} = \Omega_{\mu}^{3\,\rm int}$ is the auxiliary gauge
field introduced in Eq.\ (\ref{field tensor}). The Hopf term can be
expressed in terms of these auxiliary gauge fields $B_{\mu}$
(\ref{Hopfterm}) or equivalently in terms of the gauge fields
${\bbox \Omega}_{\mu}^{\rm int}$ as
                    \begin{eqnarray}
S_{\rm Hopf} &=& \frac{N}{96 \pi^2} \int d^2 r \,dt \,
\epsilon^{\mu \nu \lambda}\,{\bf \Omega}_{\mu}^{\rm int} \cdot 
({\bf \Omega}_{\nu}^
{\rm int}
\times {\bf \Omega}_{\lambda}^{\rm int} ).
\label{Hopf2}
                    \end{eqnarray}
The effective action for the $\bf Q^{\rm int}$ fields can now be
obtained by integrating out the spinor fields. For a slowly varying
${\bf d}$ field configuration, the ${\bf Q^{\rm int}}$ fields are
small, and it is possible to carry out a gradient expansion of the
effective action $S_{\rm eff}[{\bf Q}^{\rm int}]$ in the powers of
$\bf Q^{\rm int}$ and its derivatives. It is clear from 
Eqs.\ (\ref{gauge}) and (\ref{Hopf2}) that the Hopf
term originates from the $({\bf Q}^{\rm int})^3$ and ${\bf Q}^{\rm
int} \partial {\bf Q}^{\rm int}$ terms in the expansion of the
effective action $S_{\rm eff}[{\bf Q}^{\rm int}]$.

To calculate the effective action, we first divide the action into
two parts $ S = S_0 + S_1$, where $S_0$ and $ S_1$ are given by
                     \begin{eqnarray}
S_0 &=& \int d^2r\, dt\,
\chi^{\dagger} ({\bf r},t) \Bigg( [i \partial_0  
+\epsilon_F] I -\frac{[\hat{{\bf p}} 
- e {\bf A} ({\bf r}) ]^2 }{2 m} I
 + \gamma_0 \sigma_z \Bigg)
\chi\,({\bf r},t), \nonumber\\ 
S_1 &=& - \int d^2r\, dt\,
\chi^{\dagger} ({\bf r},t) 
\frac{1}{2} \left( \Pi^{\mu} 
Q_{\mu}^{\rm int}({\bf r},t) + Q_{\mu}^{\rm int} ({\bf r},t)
\Pi^{\mu} \right) \chi ({\bf r},t),
                        \end{eqnarray}
where $\Pi^a =  \left[ p^a - e A^a ({\bf r})\right]I/m$ and
$\Pi^0 = I $.  The effective action $S_{\rm eff}[{\bf Q}^{\rm int}]$ 
is then given by
                        \begin{eqnarray}
e^{iS_{\rm eff}[{\bf Q}^{\rm int}]} &=& \frac{\int D\chi^{\dagger} D\chi
 e^{ i ( S_0 + S_1 ) } }{\int D\chi^{\dagger} D\chi e^{ i S_0 }}
 = {\langle} e^{i S_1 } {\rangle}_{S_0}.
                        \end{eqnarray}
At this stage, we introduce a unitary transformation on the field
variables of the form
                       \begin{equation}
\chi ({\bf r},t) \rightarrow e^{\frac{i \phi_x x}{L_x}} 
e^{ \frac{i \phi_y y}{L_y}}  \chi ({\bf r},t),
\label{transformation}
                        \end{equation}
where $\phi_x$ and $\phi_y$ are constant parameters and $L_x$ and
$L_y$ are the dimensions of the system. Transformation
(\ref{transformation}) changes the boundary conditions on the single
particle wavefunctions. We shall discuss its physical meaning in more
details in the next section. At this point,
we may consider it to be a mathematical trick used to facilitate
computations. With this transformation, the action can be written as
                      \begin{eqnarray}       
S_0 &=& \int d^2r\, \frac{d\omega}{2 \pi}\,
\chi ^{\dagger} ({\bf r},\omega) \Bigg ( 
[\omega +\epsilon_F] I    
- \frac{[ {\hat{\bf p}} - e {\bf A} ({\bf r}) 
+ {\bbox{\alpha}}]^2}{2 m} I + \gamma_0 \sigma_z \Bigg ) 
\chi ({\bf r},\omega), \nonumber\\ 
S_1 &=& - \int d^2 r \, \frac{d \omega dp_0}{(2 \pi)^2}\,
 \chi^{\dagger} ({\bf r},\omega+p_0) 
\left ( \frac{1}{2} \left[ Q_{\mu}^{\rm int}({\bf r},p_o),
 \frac{\partial G_0^{-1}}
{\partial \alpha_{\mu}} \right]_+ \right) 
 \chi ({\bf r},\omega),
                        \end{eqnarray}
where $[...]_+$ means anticommutator, $\alpha^{\mu}$ =
$(\omega,\phi_x/L_x,\phi_y/L_y)$, and $G_0$ is the unperturbed Green
function. Here we have omitted the quadratic term in $Q_a$ in
$S_1$. It is easy to see that this term does not contribute to the Hopf
term in the effective action. The advantage of introducing the
parameters $\phi_x$ and $\phi_y$ also becomes clear from the
expression of $S_1$. The operators $\Pi_x$ and $\Pi_y$ can now be
conveniently expressed as the derivatives of $G_0^{-1}$ with respect to
$\phi_x$ and $\phi_y$. These parameters therefore take the place of
momenta $k_x$ and $k_y$ which are no longer good quantum numbers in a
magnetic field.

The unperturbed Green function $G_0$ is diagonal in the spin space. It
is given by
                        \begin{eqnarray}
G_0 &=& \left(
\begin{array}{cc}
G_0^+ & 0\\ 0 & G_0^-
\end{array}
\right),
\label{Green function}
                       \end{eqnarray}
where $G_0^{\pm}$ satisfy the following equations
                        \begin{eqnarray}
\left ( \omega +\epsilon_F  - 
 \frac{[ {\hat {\bf p}} - e {\bf A} ({\bf r_1}) 
 + {\bbox \alpha}]^2 }{2 m} \pm \gamma_0  \right )
 G_0^{\pm} ({\bf r_1},{\bf r_2},\omega)
 = \delta^2 ({\bf r_1}-{\bf r_2}).
\label{geq}
                        \end{eqnarray}

The perturbation expansion is now straightforward. The diagrams for the 
relevant terms in the effective action are shown in Fig.\ \ref{Graph1}. 
The first term is the familiar polarization bubble and is given by
                        \begin{eqnarray}
S_2 &=&  \frac{1}{2} \int \frac{d^2 p \,d p_0}{(2\pi)^3}
\left(\Omega_{\mu}^{i\,{\rm int}} ({\bf p},p_0) 
P^{\mu \nu} _{ij} ({\bf p},p_0) 
\Omega_{\nu}^{j\,{\rm int}}(-{\bf p},-p_0) 
\right ),
\label{second order}
                        \end{eqnarray}
\noindent
where  the tensor $P^{\mu \nu}_{ij}$ is given by
                      \begin{eqnarray}
P^{\mu \nu} _{ij} ({\bf p},p_0) &=& \frac{i}{4 L_x L_y}
{\rm Tr} \left( \int d^2 r_1 \,d^2 r_2  \frac{d \omega}{2 \pi} 
\frac{1}{2} \left[ \sigma_i e^{i {\bf p}\cdot {\bf r_1} },
\left(\frac{\partial G_0^{-1}}{\partial \alpha_{\mu}}\right)
_{{\bf r}_1} \right]_+ 
G_0 ({\bf r_1},{\bf r_2},\omega) \right. \nonumber\\
&&\left. \times 
\frac{1}{2} \left [ \sigma_j e^{-i {\bf p}\cdot {\bf r_2}}, 
\left(\frac{\partial G_0^{-1}}{\partial \alpha_{\nu}}\right)
_{{\bf r}_2} \right]_+ 
G_0 ({\bf r_2},{\bf r_1},\omega + p_0) \right ).
\label{Pi tensor}
                       \end{eqnarray}
\noindent

It is clear from the structure of $S_2$ that to get the relevant terms
of the form ${\bf Q}^{\rm int} \partial {\bf Q}^{\rm int}$, we need to
expand $P^{\mu \nu} _{ij}$ to first order in external momenta (${\bf
p},p_0$). This expansion is carried out in Appendix A. The
result is
                      \begin{eqnarray}
S_2 &=& \frac{1}{2} b^{\mu \nu \lambda}_{ij} 
 \int d^2r\, dt\, \Omega_{\nu}^{j \,{\rm int}}({\bf r},t) 
\partial_{\lambda} 
\Omega_{\mu}^{i\,{\rm int}} ({\bf r},t), \nonumber\\
 b^{\mu \nu \lambda}_{ij} &=&
\frac{1}{4 L_x L_y} {\rm Tr} \left ( \sigma_i \frac{\partial G_0^{-1}}
{\partial \alpha_{\mu}} G_0 
 \sigma_j \frac{\partial G_0^{-1}} 
 {\partial \alpha_{\nu}} G_0 
 \frac{\partial G_0^{-1}}
{\partial \alpha_{\lambda}} G_0 \right ). 
\label{h2}
                      \end{eqnarray}
\noindent

The other contribution to the Hopf term comes from the triangular
diagram shown in Fig.\ \ref{Graph1}. This diagram involves three ${\bf
Q}$ fields. 
As $ {\bf \Omega}_{\mu}^{\rm int}$ are pure gauge fields satisfying
Eq.\ (\ref{gauge}), a term with ${\bf\Omega}_{\mu}^{\rm int} \cdot (
{\bf \Omega}_{\nu}^{\rm int} \times {\bf \Omega}_{\lambda}^{\rm int}
)$ is of the same order as a $ {\bf \Omega}_{\mu}^{\rm int} (
\partial_{\lambda}{\bf \Omega}_{\nu}^{\rm int} )$ term. The
contribution of these terms come from the triangular diagram which is
given by
                    \begin{eqnarray}
S_3 &=& \frac{1}{6} \int 
\frac{d^2 p \,d^2 q \,d p_0 \,d q_0}{(2\pi)^6}\,
T^{\mu \nu \lambda}_{ijk}({\bf p},{\bf q},p_0, q_0)
 \Omega_{\mu}^{i\,{\rm int}} ({\bf p},p_0) 
\Omega_{\nu}^{j\,{\rm int}} ({\bf q},q_0) \nonumber\\
&&\times \Omega_{\lambda}^{k\,{\rm int}} 
\left(-{\bf p}-{\bf q},-p_0 - q_0\right), 
                    \end{eqnarray}
\noindent
where
                     \begin{eqnarray}
T^{\mu \nu \lambda}_{ijk}
({\bf p},{\bf q}, p_0, q_0)&=& \frac{i}{4 L_x L_y}
{\rm Tr} \left( \int d^2 r_1 \,d^2 r_2 \,d^2 r_3 \frac{d\omega}{2 \pi}  
\, \frac{1}{2} \left [ \sigma_i e^{i {\bf p}\cdot {\bf r_1}},
\left(\frac{\partial G_0^{-1}}
{\partial \alpha_{\mu}}\right)_{{\bf r}_1} \right]_+
G_0 ({\bf r_1},{\bf r_2},\omega ) 
\right. \nonumber\\
&&\left. \times 
\frac{1}{2} \left [ \sigma_j e^{i {\bf q}\cdot {\bf r_2}},
\left(\frac{\partial G_0^{-1}}
{\partial \alpha_{\nu}}\right)_{{\bf r}_2} \right]_+   
G_0 ({\bf r_2},{\bf r_3},\omega + p_0 +q_o)
\right. \nonumber\\
&& \left.  \times 
\frac{1}{2} \left [ \sigma_k e^{-i ({\bf p}+{\bf q})\cdot {\bf r_3}},
\left(\frac{\partial G_0^{-1}}
{\partial \alpha_{\lambda}}\right)_{{\bf r}_3} \right]_+ 
G_0 \left({\bf r_3},{\bf r_1},\omega +p_0 \right)\right).
                      \end{eqnarray}
For the relevant order, here we need to consider the zeroth-order term
in the expansion of $T^{\mu \nu \lambda}_{ijk}$ in powers of external
momenta. This can be done following the method outlined in Appendix A,
and one gets
                      \begin{eqnarray} 
S_3 &=& \frac{1}{6} c^{\mu \nu \lambda}_{ijk} 
\int d^2 r \,dt \,
\Omega_{\mu}^{i\,{\rm int}} ({\bf r},t) 
\Omega_{\nu}^{j\,{\rm int}} ({\bf r},t) 
\Omega_{\lambda}^{k\,{\rm int}} ({\bf r},t),  \nonumber\\ 
c^{\mu \nu \lambda}_{ijk}&=& \frac{i}{4 L_x L_y} 
{\rm Tr} \left( \sigma_i \frac{\partial G_0^{-1}}{\partial 
 \alpha_{\mu}} G_0 \sigma_j 
 \frac{\partial G_0^{-1}}{\partial
\alpha_{\nu}} G_0 
\sigma_k \frac{\partial G_0^{-1}}{\partial
 \alpha_{\lambda}} G_0 \right). 
\label{Triangle Graph}
                      \end{eqnarray}
Having obtained the relevant terms in the effective action, we now
evaluate $c^{\mu \nu \lambda}_{ijk}$ and $b^{\mu \nu \lambda}
_{ij}$. Let us first consider $ b^{\mu \nu \lambda} _{ij}$. We first
note that $G_0$ is diagonal in the spinor space, since it contains I
and $\sigma_3$. Further, it is shown in Appendix B that the
coefficients $b^{\mu \nu \lambda}_{ij}$ satisfy $b^{\mu \nu
\lambda}_{ij} = - b^{\nu \mu \lambda} _{ij}$. From these properties it
follows that all the terms with $b^{\mu \nu \lambda}_{ij}$ (\ref{h2})
for $i \ne j$ either vanishes because of trace operation in spin
indices or lead to
total derivatives which can be neglected.  Hence there are only three
non-zero terms in $b^{\mu \nu \lambda} _{ij}$.  These are
                  \begin{eqnarray}
b^{\mu \nu \lambda} _{j} &=& \frac{1}{4 L_x L_y} 
 {\rm Tr} \left(\sigma_j \frac{\partial G_0^{-1}}{\partial
\alpha_{\mu}} G_0 (\gamma_0) 
\sigma_j \frac{\partial\,G_0^{-1}}{\partial
\alpha_{\nu}}   G_0 (\gamma_0) 
 \frac{\partial G_0^{-1}}{\partial 
 \alpha_{\lambda}} G_0 (\gamma_0) \right). 
                  \end{eqnarray}
where $j$ takes values from 1 to 3, and $G_0 (\gamma_0)$ shows the
dependence of $G_0$ on the parameter $\gamma_0$. Using properties of
${\bbox \sigma}$ matrices, we get
                  \begin{eqnarray}
\sigma_{1(2)} G_0 (\gamma_0)
&=& G_0 (-\gamma_0) \sigma_{1(2)}, \nonumber\\
\sigma_{3} G_0 (\gamma_0)&=& G_0 (\gamma_0) \sigma_{3}.
                    \end{eqnarray}
Using these relations, it is now easy to eliminate the $\sigma$
matrices from the expression of $b^{\mu \nu \lambda} _{j}$. The result
is
         \begin{eqnarray}
b^{\mu \nu \lambda} _{1(2)} &=& \frac{1}{4 L_x L_y} {\rm Tr }
\left( \frac{\partial G_0^{-1}}{\partial
\alpha_{\mu}} G_0 (-\gamma_0) 
\frac{\partial G_0^{-1}}{\partial
\alpha_{\nu}}  G_0 (\gamma_0) \frac{\partial G_0^{-1}}
{\partial \alpha_{\lambda}} 
G_0 (\gamma_0) \right ),   \\
\label{cosec} 
b^{\mu \nu \lambda} _{3} &=& \frac{1}{4 L_x L_y} {\rm Tr}
\left( \frac{\partial G_0^{-1}}{\partial
\alpha_{\mu}} G_0 (\gamma_0) 
\frac{\partial G_0^{-1}}{\partial
\alpha_{\nu}}  G_0 (\gamma_0) \frac{\partial G_0^{-1}}
{\partial \alpha_{\lambda}} 
G_0 (\gamma_0) \right ). 
\label{coefficient}
        \end{eqnarray}
The contribution to the Hopf term from $S_2$ can therefore be written
as
            \begin{eqnarray}
S_2 &=& \frac{1}{2} \int d^2r\, dt\, 
\left\{ b^{\mu \nu \lambda}_{1} 
\left[ \Omega_{\nu}^{1\,{\rm int}}({\bf r},t) 
\partial_{\lambda} \Omega_{\mu}^{1\,{\rm int}} ({\bf r},t) 
+ \Omega_{\nu}^{2\,{\rm int}} ({\bf r},t) 
\partial_{\lambda}
\Omega_{\mu}^{2 \,{\rm int}}({\bf r},t) \right] 
\right. \nonumber\\
&& \left.
+ b^{\mu \nu \lambda}_{3} 
\Omega_{\nu}^{3\,{\rm int}} ({\bf r},t) 
\partial_{\lambda} \Omega_{\mu}^{3\,{\rm int}}
 ({\bf r},t) \right \}.
\label{action2}
             \end{eqnarray}

Next, let us consider $c^{\mu \nu \lambda}_{ijk}$ given by
Eq.\ (\ref{Triangle Graph}). Here we notice that
only those terms which have all ($ijk$) different produce non-zero
contributions to $S_3$. The other terms, as sketched in Appendix B, vanish
either under trace operations in spin indices or under exchange of
space-time indices 
of $c^{\mu \nu \lambda}_{ijk}$. So there are six non-vanishing terms
and a little algebra shows that their contributions are equal. 
Thus one may write
                 \begin{eqnarray}
S_3 &=&  c^{\mu \nu \lambda}_{123} \int d^2r\, dt \,
\Omega_{\mu}^{1\,{\rm int}} ({\bf r},t) 
\Omega_{\nu}^{2\,{\rm int}} ({\bf r},t) 
\Omega_{\lambda}^{3\,{\rm int}} ({\bf r},t), \nonumber\\ 
c^{\mu \nu \lambda}_{123}&=& \frac{i}{4 L_x L_y} {\rm Tr }
\left( \sigma_1 \frac{\partial G_0^{-1}}{\partial
\alpha_{\mu}} G_0 (\gamma_0) 
\sigma_2  \frac{\partial G_0^{-1}}{\partial
\alpha_{\nu}} G_0 (\gamma_0) 
\sigma_3\,\frac{\partial G_0^{-1}}
{\partial \alpha_{\lambda}} 
G_0 (\gamma_0) \right).
\label{action3}
                  \end{eqnarray}
\noindent
Using the relation $\sigma_1 \sigma_2 = i \sigma_3$ and the cyclic
property under trace operation, one can now eliminate the $\sigma$ 
matrices to get
            \begin{eqnarray}
c^{\mu \nu \lambda}_{123}
&=& - \frac{1}{4 L_x L_y} {\rm Tr }
\left( \frac{\partial G_0^{-1}}{\partial \alpha_{\mu}} 
G_0 (-\gamma_0) \frac{\partial G_0^{-1}}
{\partial \alpha_{\nu}} 
G_0 (\gamma_0) \frac{\partial G_0^{-1}}
{\partial \alpha_{\lambda}} G_0 (\gamma_0) \right) 
 = -b^{\mu \nu \lambda} _{1}, \\
\label{corela}
S_3 &=& -b^{\mu \nu \lambda} _{1} \int d^2r\, dt \,
\Omega_{\mu}^{1\,{\rm int}} ({\bf r},t) 
\Omega_{\nu}^{2\,{\rm int}} ({\bf r},t) 
\Omega_{\lambda}^{3\,{\rm int}} ({\bf r},t).
\label{threeac}
                      \end{eqnarray}

After some algebra, which is sketched in Appendix B, we find that
$S_3$ (\ref{threeac}) exactly cancels the first two terms of $S_2$
(\ref{action2}). It is shown in Appendix B that this cancelation is
independent of the explicit expressions of the coefficients $ b^{\mu
\nu \lambda}_{1(2)}$ and $c^{\mu \nu \lambda} _{123}$.  Also, it is
easy to see that the tensor $ b^{\mu \nu \lambda} _{3}$ is completely
antisymmetric in space-time indices ($\mu \nu \lambda$). As a result,
we are left with
                 \begin{eqnarray}
S_{\rm Hopf}&=& S_2 + S_3 
= \frac{1}{2} b^{\mu \nu \lambda}_{3} \int d^2r\, dt\,  
\Omega_{\mu}^{1\,{\rm int}} ({\bf r},t) 
\Omega_{\nu}^{2\,{\rm int}} ({\bf r},t)
\Omega_{\lambda}^{3\,{\rm int}} ({\bf r},t).
\label{Hopf3}
                   \end{eqnarray}
\noindent
Comparing Eqs.\ (\ref{Hopf2}) and (\ref{Hopf3}) and using
Eq.\ (\ref{coefficient}) for $ b^{\mu \nu \lambda} _{3}$, we find an
expression for the prefactor $N$ of the Hopf term:
                    \begin{eqnarray}
N &=& \frac{4 \pi}{3} 
\epsilon_{\mu \nu \lambda} b^{\mu \nu \lambda}_3
= 2 \pi {\rm Tr }
\left( \frac{\partial G_0^{-1}}{\partial
\omega} G_0 
\frac{\partial G_0^{-1}}{\partial
\phi_x} G_0 \frac{\partial G_0^{-1}}
{\partial \phi_y} G_0 \right ).
\label{integer}
                    \end{eqnarray}
where we have substituted the expression for $b^{\mu \nu
\lambda}_3$ (\ref{coefficient}) to obtain the last expression.

It is worthwhile to point out that the formal derivation of the Hopf
term in this section (\ref{Hopfterm},\ref{integer}) does
not depend crucially on the model chosen. This is manifested in the
fact that none of the cancelations of the various terms in the
effective action depend on explicit expression of the unperturbed
Green function, or equivalently on actual values of $b^{\mu \nu
\lambda}$. The symmetry properties of $b^{\mu \nu \lambda}$ required
for these cancelations are quite general, and so the result holds for
any model of type (\ref{Hamiltonian}).

\section {Evaluation of $N$}
\label{seceva}

The expression for the prefactor of the Hopf term $N$
(\ref{integer}) is essentially the same as the expression for the Hall
conductivity (\ref{Hall}). Thus we follow the work of Niu ${\it
et\,al.}$ \cite{Niu} to evaluate $N$. 

We notice that since the unperturbed Green function $G_0$ is diagonal
in spin indices, Eq.\ (\ref{integer}) can be written as
                   \begin{eqnarray}
N &=& 2 \pi  \sum_{\alpha = \pm}{\rm Tr} 
\left( \frac{\partial G_0^{\alpha-1}}{\partial
\omega} G_0^{\alpha}  \frac{\partial G_0^{{\alpha}-1}}{\partial
\phi_x} G_0^{\alpha} \frac{\partial G_0^{\alpha -1}}
{\partial \phi_y} G^{\alpha}_0 \right ).
\label{Invariant}
                   \end{eqnarray} 
The expressions for the unperturbed Green functions  $G_0^{\pm}$ can be
obtained in terms of the single-particle eigenfunctions $|n{\rangle}$
and eigenenergies $\epsilon_n^{\pm}$ as
                    \begin{eqnarray}
G_0^{\pm} &=& \sum_n |n{\rangle}{\langle}n| 
\left (\frac{\theta ( \epsilon_n^{\pm} - \epsilon_F )}
{\omega - (\epsilon_n^{\pm} - \epsilon_F) + i \eta} 
 + \frac{\theta ( \epsilon_F - \epsilon_n^{\pm})}
{\omega - (\epsilon_n^{\pm} - \epsilon_F) - i \eta } \right), 
                    \end{eqnarray}
where $\epsilon_F$ is the Fermi energy. These single-particle
eigenfunctions and eigenvalues satisfy
                  \begin{equation}
\left(\sum_a\frac{\left( {\hat p}_a + \phi_a/L_a
- e A_a({\bf r})\right)^2 }{2 m} 
\mp \gamma_0 \right) |n{\rangle} = \epsilon_n^{\pm} |n{\rangle},
\label{eigenfunction}
                   \end{equation}
where $a$ can take values x and y.  Substituting this form of the Green
function in the expression for $N$, we get
                   \begin{eqnarray}
N &=& 2 \pi i \sum_{\alpha = \pm}\sum_{n,l} 
\left( \frac{\partial G_0^{\alpha\,-1}}
{\partial \phi_x} \right)_{nl} 
\left( \frac{\partial G_0^{\alpha\,-1}}
{\partial \phi_y} \right)_{ln} \nonumber\\
&& \times 
\int\,\frac{d \omega}{2 \pi i} 
\left( \frac{\theta ( \epsilon_n^{\alpha} - \epsilon_F )}
{\omega - (\epsilon_n^{\alpha} -\epsilon_F) + i \eta}
 + \frac{\theta ( \epsilon_F - \epsilon_n^{\alpha})}
{\omega - (\epsilon_n^{\alpha} -\epsilon_F) - i \eta\,} \right ) \nonumber\\ 
&& \times \left( \frac{\theta ( \epsilon_l^{\alpha} - \epsilon_F )}
{\omega - (\epsilon_l^{\alpha} -\epsilon_F) + i \eta} 
+ \frac{\theta ( \epsilon_F - \epsilon_l^{\alpha} ) }
{\omega - (\epsilon_l^{\alpha} -\epsilon_F) - i \eta } \right )\nonumber\\
&& \times 
\left( \frac{\theta ( \epsilon_n^{\alpha} - \epsilon_F )}
{\omega - (\epsilon_n^{\alpha} -\epsilon_F) + i \eta} 
+ \frac{\theta ( \epsilon_F - \epsilon_n^{\alpha} ) }
{\omega - (\epsilon_n^{\alpha} -\epsilon_F) - i \eta } \right).
\label{nexp}
                 \end{eqnarray}

The frequency integral can now be evaluated in a straightforward
manner. Since at zero temperature only $\epsilon_0^+$ lie below the
Fermi energy, only the $\alpha = +$ term in the sum contributes. Further, 
it is important to notice that the only surviving terms in the
frequency integral are the ones where $\epsilon_n^{+}$ and
$\epsilon_l^{+}$ lie on the opposite sides of the Fermi energy
$\epsilon_F$, $\it i.e.$ the poles of the integrand lie on different
halves of the complex $\omega$ plane. Also, Eq.\ (\ref{nexp})
contains matrix elements of the momentum operator ${\bbox \Pi}_a =
\partial G_0^{-1}/\partial \alpha_a$ 
\cite{Iordanskii}. These matrix elements connect different Landau
levels, and vanish between the states of the same Landau level. This
clearly points out that it is necessary to retain the wave functions
for all Landau levels in the calculation. The use of the LLL-projected
wave functions in this case would yield an erroneous zero value for
$N$ \cite{Iordanskii,Iord}. 

After evaluating the frequency integral, the expression for the
integer $N$ becomes
                   \begin{eqnarray}
N &=& 2 \pi i  
\sum_{n \atop {\epsilon_n^+ < \epsilon_F}}
\sum_{l\atop {\epsilon_l^+ > \epsilon_F}}
 \frac{ \left[ \left( \frac{\partial G_0^{+\,-1}}
{\partial\phi_x} \right)_{nl} 
\left( \frac{\partial G_0^{+\,-1}}
{\partial \phi_y} \right)_{ln} 
- \left( \frac{\partial G_0^{+\,-1}}
{\partial\phi_x} \right)_{ln} 
\left( \frac{\partial G_0^{+\,-1}}
{\partial \phi_y} \right)_{nl} \right] }
 {(\epsilon_n^+ - \epsilon_l^+ )^2}.
                   \end{eqnarray} 
Using the relations
                    \begin{eqnarray}
\left(\frac{\partial G_0^{+\,-1}}{\partial \phi_{x(y)}}
\right)_{nl}
&=& -(\epsilon_n^+ - \epsilon_l^+ ) 
\langle n|\frac{\partial l}{\partial \phi_{x(y)}} 
\rangle, 
                     \end{eqnarray}
this can be further simplified to
                     \begin{eqnarray}
N &=&  2 \pi i \sum_{n \atop {\epsilon_n^+ < \epsilon_F}}
\left( \langle\frac{\partial n}{\partial \phi_x}| 
\frac{\partial n}{\partial \phi_y} \rangle 
- \langle \frac{\partial n}{\partial \phi_y} | \frac{\partial
n}{\partial \phi_x} \rangle \right), 
                     \end{eqnarray}
where $|\Phi_0\rangle = \prod_n |n\rangle$ is the unperturbed
many-body ground state of the system.

So far, the derivatives in the expression of $N$ are formal, and it is
not clear why $N$ has to be an invariant. To see this point more
clearly, we now consider the physical meaning of the parameters
$\phi_x$ and $\phi_y$. The ground state of the system in this case is
a Slater determinant of the single-particle states, which are
solutions to Eq.\ (\ref{eigenfunction}). These states are constructed
by Niu $\it{et.al.}$ in a slightly different context \cite{Niu}, and
are given by
               \begin{eqnarray}
\psi_n (x,y) &=& e^{\frac{-i \phi_x x}{L_x}}
e^{\frac{-i \phi_y y}{L_y}} W_n (x,y) ,\nonumber\\ 
W_n (x,y)&=& \sum_{m = -\infty}^{\infty} 
e^{ i \Lambda_n \beta_m l_B^2 }
e^{i \beta_m y } u_0 ( x - \beta_m l_B^2 ),
\label{wavefunction}
                 \end{eqnarray}
where $\Lambda_n = (2 \pi n + \phi_x)/L_x$, $\beta_m = (2\pi m 
+ \phi_y)/L_y$ and $u_0$ is the ground state wave function for
a harmonic oscillator. From this, it is easy to check that $\psi_n$ is
the solution to the Hamiltonian H with eigenvalues
$\epsilon_0^+$. Furthermore, $W_n(x,y)$ also satisfies the boundary 
conditions
                  \begin{eqnarray}
W_n( x + L_x , y )&=& e^{ i\phi_x } 
e^{i y L_x/l_B^2} W_n (x,y),  \nonumber\\
W_n ( x, y + L_y ) &=& e^{ i \phi_y} W_n(x,y),   
\label{boundary condition}
                  \end{eqnarray}
where $L_x, L_y$ are the dimension of the system.

From these relations, it is clear that we can interpret $\phi_x$ and
$\phi_y$ as the boundary phase parameters of the system. So, variation
of these parameters means variation of the boundary conditions for the
ground-state wavefunction. Following Niu $\it{et\,al.}$ \cite{Niu}, we
now argue that the value of the invariant $N$ is independent of the
boundary condition chosen, so that we can replace this expression by
its average over all possible boundary conditions. This allows us to
write
                 \begin{eqnarray}
N &=&\frac{1}{2\pi i} \int_0^{2 \pi}\ d \phi_x 
\int_0^{2 \pi} d \phi_y  
\left( \langle \frac{\partial \Phi_0}{\partial \phi_y}| 
\frac{\partial \Phi_0}{\partial \phi_x} \rangle 
 - \langle \frac{\partial \Phi_0}{\partial \phi_x} | 
\frac{\partial \Phi_0} {\partial \phi_y} \rangle \right). 
                  \end{eqnarray}
It is easy to see from the condition (\ref{boundary condition})
on the single-particle wavefunctions that $(2\pi ,2\pi)$ and (0,0)
are same point in the $\phi$ space. As a result, we can interpret $N$
as a surface integral over a closed surface in the parameter space,
which can be expressed as a line integral. This line integral is
              \begin{eqnarray}
N &=& \frac{1}{2 \pi i} \oint {\bf d}{\vec \phi } \cdot 
\langle \Phi_0 | {\bbox \nabla_{\vec \phi}} \Phi_0 \rangle. 
\label{line integral}
               \end{eqnarray}

Since $\phi_x$ and $\phi_y$ are parameters in the Hamiltonian, the
line integral in Eq.\ (\ref{line integral}) can be interpreted as the
Berry's phase \cite{Berry} picked up by the ground-state wavefunction
as it moves around a closed contour in the parameter space. But since
in this case, the ground state is separated by a finite energy gap from
the excited states and is non-degenerate, the many-body ground-state
wavefunction must return to itself after traversing the contour. So,
the value of the line integral must be 2$\pi$i times an integer, which
immediately tells us that $N$ must be an integer. The value of this
integer can be explicitly evaluated by constructing the ground-state
wavefunction as a Slater determinant of the single-particle states
(\ref{wavefunction}) and by choosing a rectangular contour in the
parameter space. The calculation is straightforward
\cite{Niu,Kohomoto} and for $\nu = 1$, it yields the value $N = 1$.

From the above discussion, it is clear that the above result is not an
artifact of the simple model chosen to describe the system.  The
properties of the many-body ground state that we used to argue that
$N$ should be an integer are the presence of a finite gap between the
ground state and the other excited states, and the non-degeneracy of
the ground-state wavefunction. As long as these conditions are
satisfied, the value of invariant $N$, which can not be a continuous
function of the model parameters, must have the same value as that
obtained from the calculation based on this simple model. It can
therefore be argued that this result is robust against presence of
weak disorder ($\omega_c \tau \gg 1$) in the system, which broadens
the Landau levels to bands but do not lead to mixing of different
Landau levels. A more mathematically rigorous justification of this
issue is given by Niu ${\it et\,al.}$ \cite{Niu} and Ishikawa ${\it
et\,al.}$ in \cite{ishikawa} in the context of Hall conductivity in
quantum Hall systems.

\section  {Spin, Charge, and Gauge Invariance}
\label{secspi}

In this section, we compute the spin and charge densities for the
skyrmion, as well as its total charge and spin. To do this, for
reasons that will become clear later, we first consider the gauge
invariance of our effective action. We have mentioned before that the
$B$ fields introduced in Eq.\ (\ref{field tensor}) are auxiliary gauge
fields. The gauge transformation
here corresponds to an arbitrary space-time-dependent rotation of the
spin-quantization reference frame about
the local ${\bf d}$ axis, since such a rotation does not change the
physical state of the system. It can be easily seen that under such a
transformation $U_{\Lambda} = e^{
\frac{i}{2}({\bbox \sigma}\cdot {\bf d}) \Lambda(x,y,t) }$, the
auxiliary field $B_{\mu}$ transform as
              \begin{equation}
B_{\mu} \rightarrow B_{\mu} + \partial_{\mu} \Lambda. 
\label{gaugetrans}
              \end{equation}
\noindent
The action (\ref{Hopfterm}), however, is not invariant under this
transformation. It acquires an additional surface term
                 \begin{eqnarray}
\delta S &=&  -\frac{N}{16 \pi} \oint dl_a \,dt\,
(B_{a} \partial_{0} \Lambda  - B_{0} \partial_{a} \Lambda),
                  \end{eqnarray}
where $dl_a$ is the length element along the edge, and $\oint$ denotes
integration along the edge. From now on, we shall assume a rectangular
boundary for simplicity.

Such a non-invariance of the bulk action under gauge transformation is
well known for both Abelian and non-Abelian gauge theories with
Chern-Simons term in non-compact space \cite{Wilczek1}. The most
well-known example for the Abelian case is the effective action for
integer and fractional quantum Hall states: $ S_{\rm eff} =
(\sigma_{xy}/4) \int d^2r\,dt \,\epsilon^{\mu \nu \lambda} A_{\mu}^{\rm
em} F_{\nu \lambda}^{\rm em}$. This non-invariance is merely a
statement that in a space with boundary the bulk effective action is not the
complete action for the problem, and we must add the edge action
$S_{\rm edge}$, so that the total action $ S_{\rm total} = S_{\rm
bulk} + S_{\rm edge}$ is gauge invariant. In case of quantum Hall
states, such a consideration leads to gapless edge excitations
\cite{Wen}. As we shall see, in our case, considering the edge
action is absolutely crucial for determining the correct value of the
contribution of the Hopf term to the total spin of the skyrmion.

The 1+1D edge effective action that we need to construct, must
transform under the gauge transformation (\ref{gaugetrans}) in
such a manner so as to cancel the gauge non-invariant term of the bulk
action:
                       \begin{equation}
\delta S_{\rm edge} =  \frac{N}{16 \pi} \oint dl_a \,dt\,
( B_{a} \partial_{0} \Lambda - B_{0}\partial_{a} \Lambda).
                        \end{equation}
\noindent
The simplest edge action that achieves this can be written in terms of
a boson field $\eta$ representing the edge excitations
\cite{Baez,Wilczek1}:
\begin{eqnarray}
S_{\rm edge} &=&  \oint dl_a \,dt\, \left[ (D_{t} \eta)^2 
- \kappa (D_{a} \eta)^2 \right]  + \frac{N}{16 \pi} 
\oint dl_a \,dt  \,
( B_{a}\partial_{0} \eta  -  B_{0}\partial_{a} \eta ).
\label{eac}
\end{eqnarray}
In Eq.\ (\ref{eac}), $D_{\mu}$ is the covariant derivative defined as
$D_{\mu} = (\partial_{\mu} - B_{\mu}) $, $\kappa$ is a real parameter,
and the boson field $\eta$ 
transforms under the gauge transformation (\ref{gaugetrans}) as
$\eta \rightarrow \eta + \Lambda$. It is easy to check that under
these conditions the first term of the edge action is invariant under
gauge transformation, while the second term produces the correct extra
term $\delta S_{\rm edge}$, which cancels gauge contribution of the
bulk action, so that the total action $S_{\rm total}$ remains gauge
invariant.

Next, we proceed to compute the spin density of the skyrmion. The
easiest way to do this is to consider the response of the system to
arbitrary rotation by an infinitesimal solid angle ${\bbox
\theta}({\bf r},t)$. The action after such a rotation can be written
as
                      \begin{eqnarray}
S[\psi,\psi^{\dagger},{\bf d},Q_{\mu}^{\bbox \theta}] 
&=& \int d^2r \,dt\,
\psi^{\dagger}_a({\bf r},t) \Bigg ( [i\partial_0 
+\epsilon_F]I - Q_0^{\bbox \theta} \nonumber\\
&& - \frac{\left \{[{\hat {\bf p}} - e{\bf A}({\bf r})]I 
- {\bf Q}^{\bbox \theta}({\bf r},t)\right \}^2}{2m} 
 + \gamma_0 {\bbox \sigma}\cdot{\bf d} \Bigg )_{ab}
\psi_b ({\bf r},t),
                       \end{eqnarray}
\noindent
where $Q_{\mu}^{\bbox \theta} = -i U_{\bbox \theta}^{-1}
(\partial_{\mu} U_{\bbox \theta}) = {\bbox \sigma}\cdot
\partial_{\mu}{\bbox \theta}({\bf r},t)/2$ and $U_{\bbox \theta} =
e^{\frac{i}{2} {\bbox \sigma}\cdot {\bbox \theta}({\bf r},t)}$. So, a
variation of the effective action with respect to $\partial_0 {\bbox
\theta}({\bf r},t)$ gives the spin ${\bf J}$ of the skyrmion:
                 \begin{eqnarray}
{\bf J} &=& \int d^2r\, dt\, \frac{\delta S_{\rm eff}}
{\delta [\partial_0 {\bbox \theta}({\bf r},t)]}.
                   \end{eqnarray}

To compute the effective action, we again rotate the inhomogeneous
${\bf d}$ field to a homogeneous configuration ($\bf d = {\bf e_z}$)
and proceed in the same manner as the previous section. The difference
is that now the internal gauge field $Q_{\mu}^{\rm int}$ is replaced
by the total gauge field $Q_{\mu}^{\rm total} = Q_{\mu}^{\rm int} +
U^{-1} Q_{\mu}^{\bbox \theta}U$. The details of the computations are
essentially the same as in the previous section. Specially, the
cancelation of the different terms are exactly the same since the field
tensor $f_{\mu \nu}$ (\ref{field tensor}) corresponding to
$Q_{\mu}^{\rm total}$ is zero. After some 
algebra, it can be seen that the form of both the edge and the bulk
effective actions remain the same, with the auxiliary fields $B_{\mu}$
and the scalar field $\eta$ replaced by
                    \begin{eqnarray}
B_{\mu} \rightarrow B_{\mu} + {\bf d}\cdot \partial_{\mu}
{\bbox \theta},\nonumber\\
F_{\mu \nu} \rightarrow F_{\mu \nu} + \partial_{\mu}{\bf d}\cdot 
\partial_{\nu}{\bbox \theta} - \partial_{\nu}{\bf d}\cdot 
\partial_{\mu}{\bbox \theta}, \nonumber\\ 
\eta \rightarrow \eta + {\bf d}(\infty)\cdot {\bbox \theta},
                    \end{eqnarray}
where ${\bf d}(\infty) = {\bf e_z}$ is the constant unit vector at
the edge.  The total action $S_{\rm total}$ now is the sum of the bulk
and the edge actions. There are two terms contributing to the
spin of the skyrmion in the bulk. The first term is
linear in $B$ fields and has the same origin as the Berry term in
Eq.\ (\ref{ferrob}), while the second term is the Hopf term. The
relevant part of the effective action is 
                  \begin{eqnarray}
S_{\rm bulk} &=&  \frac{\rho_0}{2}\int d^2r\, dt\,( B_{0} 
+ {\bf d} \cdot \partial_{0}  {\bbox \theta} )
+\frac{N}{32 \pi}\int d^2r\, dt \,
\epsilon^{\mu \nu \lambda} \left (B_{\mu} 
+ {\bf d} \cdot \partial_{\mu} {\bbox \theta} \right) 
\left(F_{\nu \lambda} + 2 \partial_{\nu}{\bf d}\cdot 
\partial_{\lambda}{\bbox \theta} \right ),\nonumber\\
S_{\rm edge} &=& \frac{N}{16 \pi} \oint dl_a\, dt\, 
\left\{ \left [\partial_{0} \eta + {\bf d}(\infty) \cdot 
\partial_{0} {\bbox \theta} \right ] \left [B_{a} 
+{\bf d}(\infty) \cdot \partial_{a} {\bbox \theta} \right ]
\right.\nonumber\\
&& \left. 
-\left [\partial_{a} \eta + {\bf d}(\infty) \cdot 
\partial_{a} {\bbox \theta} \right ] \left [B_{0} 
+{\bf d}(\infty) \cdot \partial_{0} {\bbox \theta} \right]
\right \}.
\label{spinaction}
\end{eqnarray}
\noindent
The variation of Eq.\ (\ref{spinaction}) with respect to
$\partial_0{\bbox \theta}({\bf r},t)$ gives the spin of the skyrmion
relative to the ferromagnetic ground state:
             \begin{mathletters}
\begin{equation}
{\bf J} ={\bf J}^{{\rm Berry}} + {\bf J}^{{\rm Hopf}} 
+ {\bf J}^{\rm edge},
\label{spina}
\end{equation}
\begin{equation}
{\bf J}^{{\rm Berry}} = \frac{\rho_0}{2}
\int d^2r\, ({\bf d}-{\bf e}_z), 
\label{spinb}
\end{equation}
\begin{equation}
{\bf J}^{{\rm Hopf}} = \frac{N}{16 \pi} \int d^2r\, 
\epsilon^{ab0} \left[\partial_a {\bf d} \times 
\partial_b {\bf d}  + \partial_a ( {\bf d} B_b) \right], 
\label{spinc}
\end{equation}
\begin{equation}
{\bf J}^{\rm edge} = \frac{N {\bf d}(\infty)}{16 \pi} \oint dl_a \, 
\epsilon^{ab0} B_{b}. 
\label{spind}
\end{equation}
             \end{mathletters}
\noindent
The first term in Eq.\ (\ref{spina}) comes from the Berry term in the
action (\ref{spinaction}), while the second and the third terms are
contributions to the skyrmion spin from the Hopf and the edge
terms. We first analyze the latter contribution.  The first term in
the integrand of ${\bf J}^{\rm Hopf}$ (\ref{spinc}) is the
contribution to the local spin density of the skyrmion from the Hopf
term and is given by ${\bf j}_0({\bf r},t) = (N/16 \pi) \epsilon^{ab0}
\partial_a {\bf d} \times \partial_b {\bf d}$. This local spin density
is directed along the local ${\bf d}$ vector and is a total
derivative, so it produces no net contribution to the total spin of
the skyrmion. The net contribution of the Hopf term to the total spin
therefore comes from the second term in ${\bf J}^{\rm Hopf}$ and from
${\bf J}^{\rm edge}$ (\ref{spind}).  Using the expression for the
topological charge $Q_{\rm top} = (1/4 \pi) \int d^2r\, {\bf d}\cdot (
\partial_x {\bf d} \times \partial_y {\bf d} )$ and Eq.\ (\ref{field
tensor}), it is easy to see that each of these terms contribute
equally to the total skyrmion spin. The total contribution to the
skyrmion spin from the Hopf term is therefore $ N{\bf d}(\infty)Q_{\rm
top}/2$. It is directed along ${\bf d}(\infty) = {\bf e}_z$ and for a
skyrmion with unit topological charge $Q_{\rm top}=1$ has the
magnitude of $1/2$ at $\nu = 1$.  Since the last term in the
expression of the spin density comes from $S_{\rm edge}$, we see that
it is crucial to include the edge action in the calculation of the
contribution of the Hopf term to the total spin. Omitting the edge
action would give only half of the actual result.

Next, we consider the contribution to the skyrmion spin from the Berry
term (\ref{spinb}). This contribution is clearly proportional to the
number of flipped spin in the skyrmionic state, and increases with the
size of the skyrmion \cite{Moon}. So in the case of a long-wavelength
skyrmion, the Berry term produces dominant contribution to the
skyrmion spin. The exact value of this contribution, unlike the
contribution due to the Hopf term, is not a universal number, but
depends on the details of the field configuration. So we conclude that
the spin of the skyrmion in quantum Hall ferromagnets, as measured as
a response to an external field, is not quantized to integral or
fractional multiples of $1/2$. This is in contrast to the case of
skyrmions in antiferromagnetic systems, which have a universal value of
the spin completely determined by the coefficient of the Hopf term.

Finally, we compute the charge density and the total charge of the
skyrmion. To do this, we compute the response of the system to an
external electromagnetic potential $A_{\mu}^{\rm em}$. The action, in
presence of an external electromagnetic potential can be
written as
                    \begin{eqnarray}
S[\psi,\psi^{\dagger},{\bf d},A_{\mu}^{\rm em}] &=& \int d^2r\, dt\,
\psi^{\dagger}_a({\bf r},t) \Bigg ( \left[i\partial_0  
- e A_0^{\rm em} ({\bf r},t) +\epsilon_F \right] I  \nonumber\\
&& - \frac{\left[{\hat {\bf p}} - e{\bf A}({\bf r}) 
- e{\bf A}^{\rm em}({\bf r},t)\right]^2  }{2m} I 
 + \gamma_0 {\bbox \sigma}\cdot{\bf d} \Bigg )_{ab}
\psi_b ({\bf r},t),
                     \end{eqnarray}
\noindent
and the charge density $\rho$ of the skyrmion can be obtained by
varying the effective action, with respect to $A_0^{\rm em}$.

To calculate the effective action, we follow the procedure of Sec.\
\ref{secexp} noting that $A_{\mu}^{\rm em}$ is a U(1) gauge field
added to the SU(2) gauge field ${\bf Q}_{\mu}^{\rm int}$. Taking into
account that the unperturbed Green function $G_0$ is diagonal in spin
space, we find that any term in $S_2$ of the form $A_{\mu}^{\rm
em} \partial_{\nu} \Omega_{\lambda}^{1(2)\,{\rm int}}$ vanishes due to
the trace operation in spin indices. Similarly, it can be shown that all terms
involving product of $A_{\mu}^{\rm em}$ and $\Omega^{\rm int}$ in
$S_3$ also vanishes. So the cancelation of the terms in $S_2$ and
$S_3$ follows exactly in the same way as in Sec.{\ref{secexp}, and
finally we are left with the effective action
                   \begin{eqnarray}
S_{\rm eff} [B_{\mu},A_{\mu}^{\rm em}] &=& 
\frac{N}{32 \pi} \int d^2r\, dt \,
\epsilon^{\mu \nu \lambda} B_{\mu} F_{\nu \lambda}  
 + \frac{N_1 e}{8 \pi} \int d^2 r\, dt \,
\epsilon^{\mu \nu \lambda} B_{\mu} F_{\nu \lambda}^{\rm em}  
\nonumber\\
&& + \frac{N e^2}{8\pi}\int d^2r \,dt 
\epsilon^{\mu \nu \lambda} 
A_{\mu}^{\rm em} F_{\nu \lambda}^{\rm em},
                   \end{eqnarray}
\noindent
where the coefficient $N_1$ is given by
                    \begin{eqnarray}
N_1 &=& 2 \pi {\rm Tr}
\left(\sigma_3 \frac{\partial G_0^{-1}}{\partial \omega} 
G_0 \frac{\partial G_0^{-1}}{\partial \phi_x} G_0
\frac{\partial G_0^{-1}}{\partial \phi_y} G_0 \right).
                     \end{eqnarray}
Since the unperturbed ground state is completely spin polarized, it
can be easily seen that $N_1= N$. The variation with respect to
$A_0^{\rm em}$ now gives the charge density $\rho$
                       \begin{eqnarray}
\rho ({\bf r},t) &=& \left(\frac{\delta S_{\rm eff}}
{\delta  A_{0}^{\rm em}} \right)_{A_{\mu}^{\rm em} = 0}
= \frac{N e}{4 \pi} {\bf d} 
\cdot ( \partial_x {\bf d} \times \partial_y {\bf d}).
\label{cd}
                       \end{eqnarray}
The electric charge density (\ref{cd}) is found to be proportional to the
topological charge density. The space integral of this charge density
gives the total charge of the skyrmion, given by 
                      \begin{eqnarray}
Q_{\rm em} &=& \frac{N e}{4 \pi} \int d^2r {\bf d} 
\cdot ( \partial_x {\bf d} \times \partial_y {\bf d}) = N e Q_{\rm top}.
                       \end{eqnarray}
For $\nu = 1$, the skyrmion with $Q_{\rm top} = 1$ therefore carries a
charge $e$. For fractional filling, as we shall see in Sec. \ref{secfra}, the
prefactor $N$ has fractional value leading to fractional charge for
the skyrmion.

\section{Statistics}
\label{secstat}

In this section, we outline the argument of Wilczek and Zee
\cite{Wilczek} to obtain spin, charge and statistics of skyrmions in
quantum Hall ferromagnets. For this purpose, let us recall that the
effective action of a skyrmion in quantum Hall ferromagnets is given
by Eqs.\ (\ref{ferroa}) and (\ref{Hopfterm}). First, we compute the $z$
component of the spin of the skyrmion. For this purpose, we introduce
an adiabatic time evolution which corresponds to a rotation of the
${\bf d}$ field in the XY plane about the center of the skyrmion, such
that the wavefunction returns to itself after time T. The phase
acquired by the skyrmion wavefunction $\psi$ in the process is
\begin{eqnarray}
\psi({\bf r},T) &=& e^{i\int_0^T dt \int d^2r ( L_{\rm Berry} 
+ L_{\rm E} + L_{\rm Hopf} )} \psi({\bf r},0),
\label{phase1}
\end{eqnarray}
where $ L_{\rm Berry} = \rho_0 B_0 /2$, $ L_E = - \kappa^2 \left(
{\nabla} {\bf d} \right)^2 $ and $L_{\rm Hopf} = N/32\pi \,
\epsilon^{\mu \nu \lambda} B_{\mu} F_{\nu \lambda}$ are the
Lagrangians corresponding to $S_{\rm FM}$ (\ref{ferroa}) and $S_{\rm
Hopf}$ ({\ref{Hopfterm}).  But such a time evolution also corresponds
to rotation of the skyrmion spin by $2\pi$ around the z axis in the
spin-space \cite{Wilczek,Volovik4}.  So the phase picked up by the
wavefunction in this process must be $\exp(2\pi i J_z)$, where $J_z$
is the $z$ component of the skyrmion spin. So $J_z$ must be given by
\begin{eqnarray}
J_z &=& \frac{1}{2\pi} \int_0^T dt \int d^2r ( L_{\rm Berry} 
+ L_{\rm E} + L_{\rm Hopf} ).
\label{skspin1}
\end{eqnarray}
To compute the phase picked up by the skyrmion wave-function, we now
make a specific choice for the ${\bf d}$ field configuration
\cite{Volovik4}:
\begin{eqnarray}
{\bf d}({\bf r},t) &=& \cos[\gamma(|{\bf r}|)]{\bf e}_z
+ \sin[\gamma(|{\bf r}|)] \{ \cos[\phi +\phi_0(t)]{\bf e}_{r} + \sin[\phi + 
\phi_0(t)]
{\bf e}_{\phi} \},
\label{dfield}
\end{eqnarray}
where $\phi$ is the azimuthal angle, $ \cos[\gamma(|{\bf r}|)] = (r^2
-\Lambda^2) /(r^2 +\Lambda^2)$, $\Lambda$ is a parameter depending on
the skyrmion radius and $\phi(t)$ is chosen to be $2\pi t/T$ to ensure
that the ${\bf d}$ field configuration returns to itself after a time
T.  With this choice of ${\bf d}$ field configuration, $Q_{\rm top} =
1$ and the fields $B_{\mu}$ can be represented as 
\begin{eqnarray}
B_{\mu} &=& \{1 + \cos[\gamma(|{\bf
r}|)]\} \partial_{\mu} (\phi +\phi_0(t)).
\label{bfield1}
\end{eqnarray}

Using Eqs.\ (\ref{phase1}) and (\ref{bfield1}), it can
be seen that the phase picked up by the skyrmion has three distinct
contribution. The contribution from the second term $L_E$ gives the
usual factor $\exp(iET)$, where $E$ is the skyrmion energy. The other
two contributions to the phase comes from $L_{\rm Berry}$ and
$L_{\rm Hopf}$ which are given by
\begin{eqnarray}
\theta_{\rm Berry}^s &=& 2 \pi \frac{\rho_0}{2} 
\int d^2 r \, \{ 1 + \cos[\gamma(|{\bf r}|)]  \}, \nonumber\\
\theta_{\rm Hopf}^s &=& 2 \pi \frac{N}{8 \pi}\int d^2 r \, \{1 +
\cos[\gamma(|{\bf r}|)]\} F_{xy}({\bf r}),
\label{phase2}
\end{eqnarray}
where $F_{xy}({\bf r}) = \nabla \times {\bf B}({\bf r})$. From
Eqs.\ (\ref{phase1}), (\ref{skspin1}) and (\ref{phase2}), we get 
the expression for the $z$ component of the spin of the skyrmion, 
over and above the ferromagnetic background, to be
\begin{eqnarray}
J_z &=& \frac{\rho_0}{2} 
\int d^2 r \, \{\cos[\gamma(|{\bf r}|)] -1 \} 
+ \frac{N}{8 \pi}\int d^2 r \, 
\{1 +\cos[\gamma(|{\bf r}|)]\} F_{xy}({\bf r}).
\label{skspin2}
\end{eqnarray}

The first term in the expression for the spin, represents the
contribution of the Berry term to the skyrmion spin and its
contribution can be seen to be same as Eq.\ (\ref{spinb}). The second term
comes from the Hopf term in the action. The contribution of the second
term can be directly evaluated by computing $F_{xy}$ using the expression of
$B_{\mu}$ in Eq.\ (\ref{bfield1}). After some algebra, one obtain this 
contribution to be $N/2$ and so the spin is given by
\begin{eqnarray}
J_z &=& \frac{\rho_0}{2} \int d^2 r \, 
\left[({\bf d}-{\bf e}_z) \cdot {\bf e}_z \right] + \frac{N}{2},
\end{eqnarray}
which agrees to the expression of the spin obtained in the last
section. It is to be noted that the Berry term in the effective action
of ferromagnets is linear in time derivative. So its contribution to
the phase and hence to the skyrmion spin does not vanish in the
adiabatic limit. In contrast, the effective action of antiferromagnets
\cite{Zee,Haldane} has a term quadratic in time derivative. The
contribution of this term vanishes as $1/T$ in the adiabatic limit
\cite{Wilczek}, and hence the only contribution to the spin of the
skyrmions in antiferromagnetic systems comes from the Hopf term.

Next, we compute the charge of the skyrmion. To do this, we rotate the
skyrmion adiabatically around a circle of radius $R_0$ in the XY plane
and compute the phase picked up by the skyrmion. Since the skyrmion is
a charged object and we have a magnetic field ${\bf H}_0$
perpendicular to the plane, the phased picked up by the skyrmion
wavefunction must be given by $\exp(ie^{*}\Phi/\Phi_0)$, where $\Phi$
is the magnetic flux through the circle, $e^{*}$ is the skyrmion
charge, and $\Phi_0 = \pi/e$ is the flux quantum. On the other hand,
this rotation can again be thought as an adiabatic time evolution of
the skyrmion wavefunction. So using Eq.\ (\ref{phase1}), we can obtain
an expression for the charge of the skyrmion to be
\begin{eqnarray}
e^{*} &=& \frac{\Phi_0}{\Phi} \int_0^T dt \int d^2r ( L_{\rm Berry} 
+ L_{\rm Hopf} ),
\label{skcharge1}
\end{eqnarray}
where ${\bf r}$ denotes coordinates as measured from the center of the
circle.

To evaluate the integral in Eq.\ (\ref{skcharge1}), we first rewrite the
expression of the $B_{\mu}$ fields in terms of complex coordinates as
\begin{eqnarray}
B_{\mu} &=& \{1 + \cos[\gamma(|z -Z_0(t)|)]\} \partial_{\mu}
\left[ \ln (z -Z_0(t)) - \ln (|z -Z_0(t)|) \right]. 
\end{eqnarray}
where $z = x +iy $, $Z(t) = X_0 + iY_0$ represent the coordinates of the
center of the skyrmion and the time evolution is such that the
skyrmion rotates once around the circle in time $T$, ${\it i.e.}$
$Z(T) = Z(0)$. Then the contribution of the Berry term to the phase is 
\begin{eqnarray}
\theta_{\rm Berry}^c &=& \frac{\rho_0}{2}\int d^2z  
\int_0^T \{1 + \cos[\gamma(|z
-Z_0(t)|)]\} \partial_{t} \ln (z -Z_0(t)) \nonumber\\
&=& \rho_0i \int d^2z \oint dZ_0 \frac{1}{z -Z_0(t)}
+ \frac{\rho_0}{2} i\int d^2z  \int_0^T \{ \cos[\gamma(|z
-Z_0(t)|)]-1\} \frac{ \partial_{t}Z_0(t)}{z -Z_0(t)}.
\label{skcharge2}
\end{eqnarray}
The evaluation of the first term on the right-hand side of Eq.\
(\ref{skcharge2}) is straightforward and gives $2 \pi^2
R_0^2\rho_0$. To evaluate the second term, we interchange the order of
integration, and shift the integration variable to $z' = z -Z_0$. Then
the space integral can be written as $\int
d|z'|\{\cos[\gamma(|z'|)]-1\} \int d\phi' \exp(-i\phi')$. Since the
integrand $\{\cos[\gamma(|z'|)]-1\}$ vanishes at infinity, the $z'$
integral converges, and hence integration over $\phi'$ yields a zero
result. So, the net contribution from the Berry term to the charge
comes from the first term. The contribution from the Hopf term to the
phase can be written as
\begin{eqnarray}
\theta_{\rm Hopf}^c &=& \frac{N}{16 \pi}
\int d^2z  \int_0^T \{1 + \cos[\gamma(|z
-Z_0(t)|)]\} F_{xy}(|z -Z_0(t)|)
\frac{1}{z -Z_0(t)} \partial_{t}Z_0(t). 
\end{eqnarray}
Since $F_{xy}$ is a localized function which vanishes at infinity, it
can be shown, following similar logic as in the case of the Berry
term, that the contribution of the Hopf term to the phase vanishes. So
we finally get 
\begin{eqnarray}
e^{*} &=& \frac{\Phi_0}{\Phi} 2\pi^2 R_0^2 \rho_0
= Ne
\end{eqnarray}
where in the last step we have used the relation $\rho_0 = N/2 \pi
l_B^2$. This result agrees to that of Yang and Sondhi \cite{Yang},
obtained by explicitly computing the Berry phase using a variational
wavefunction. 

It is to be noted, that the contribution to the charge of the
skyrmion, within this prescription, comes from the Berry term in the
action. On the other hand, when we calculated the charge of the
skyrmion as a response of the effective action to an external
electromagnetic field in Sec.\ref{secspi}, we found that the
charge of the skyrmion is a consequence of the coupling of the Hopf
term to the external electromagnetic field. This apparent
contradiction can be resolved in the following way. Skyrmions in
quantum Hall ferromagnets have a non-zero electric
charge and consequently couple to electromagnetic fields. Now, let us
consider the effective action of a 
skyrmion in presence of a constant magnetic field $H_0$ in the z
direction. The coupling of the Hopf term in the effective action to
this field would give us a term
\begin{eqnarray}
S_{\rm em} &=& \frac{N e}{8 \pi} \int d^2 r\, dt \,
\epsilon^{\mu \nu \lambda} B_{\mu} 
F_{\nu \lambda}^{\rm em},\nonumber\\ 
&=& \frac{N e H_0}{4 \pi} \int d^2 r\, dt \, B_0 
= \frac{\rho_0}{2}  \int d^2 r\, dt \, B_0  = S_{\rm Berry}
\end{eqnarray}
So, we see that the Berry term in the effective action of skyrmions in
quantum Hall ferromagnets is a consequence of the non-zero electric charge
of the skyrmions and the presence of a constant magnetic field $H_0$
in the system. Therefore it is not surprising that the contribution to
the charge, obtained by rotating the skyrmion, comes from the Berry
term. It is to be noted that in antiferromagnetic system like ${\rm
He}^3$-A films \cite{Volovik1}, the skyrmions are electrically neutral
and hence we do not have a Berry term in their effective action.

Finally, we compute the statistical phase picked up due to exchange of
two skyrmions. To calculate this statistical phase, the first skyrmion
is adiabatically rotated around a circle of radius $R_0$ with the
second skyrmion at the center. The statistical phase is then defined
as the ${\it additional}$ phase picked up by the wave-function of the 
first skyrmion due to
presence of the second skyrmion at the center.  The radius of the
circle is chosen to be much larger than radii of both the skyrmions,
so that these skyrmions can be assumed to be non-interacting. In this
case, the total ${\bf d}$ field due to the skyrmions are given by
${\bf d} = {\bf d}^1+{\bf d}^2-{\bf z}$, where superscripts $1$ and
$2$ refer to the first and the second skyrmion respectively.
Consequently the Lagrangian of the two-skyrmion system can be shown to
be given by
\begin{eqnarray}
L &=& L_{\rm Berry} + L_{\rm Hopf}, \nonumber\\
L_{\rm Berry}&=& \frac{\rho_0}{2} [B_0^1({\bf r}- {\bf R}_0) +
B_0^2({\bf r})], \nonumber\\
L_{\rm Hopf} &=&  \frac{N}{32\pi} \epsilon^{\mu \nu \lambda}
[B_{\mu}^1({\bf r}- {\bf R}_0) + B_{\mu}^2({\bf r}) ]
[F_{\nu \lambda}^1({\bf r}- {\bf R}_0) + F_{\nu \lambda}^2({\bf r})].
\end{eqnarray} 
From the Lagrangian, it is clear that the additional phase appears due
to the cross term $N/32\pi \, \epsilon^{\mu \nu \lambda}
B_{\mu}^1({\bf r}- {\bf R}_0)F_{\nu \lambda}^2({\bf r})$ in the
Lagrangian. The Berry term, being linear in
$B$ fields, do not contribute to the statistical phase.  The
contribution to the statistical phase therefore comes from the Hopf
term and is given by
\begin{eqnarray}
\theta_{\rm stat} &=&  \frac{N i }{16\pi}\int d^2z  
\{1 + \cos[\gamma(|z-Z_0(t)|)]\} F_{xy}^2(z)
\oint \frac{1}{z -Z_0(t)}. 
\label{statint}
\end{eqnarray}
Since the integrand in Eq.\ (\ref{statint}) is non-zero only within
the radius of the second skyrmion at the center, where $\{1 +
\cos[\gamma(|z-Z_0(t)|)]\}$ is a constant, the contour integral can be
easily evaluated and we finally get $\theta_{\rm stat} = N \pi$. The
statistics of the skyrmion is, therefore, completely determined by the
coefficient of the Hopf term. Consequently, the skyrmions would be
Fermions for odd integer fillings and, as we shall see in the next
section, anyons for fractional filling factors $\nu = 1/(2s+1)$. Our
result for the skyrmion statistics here agrees to that of Yang and
Sondhi \cite{Yang}, but contradicts that of Dziarmaga
\cite{Dziarmaga}.

The spin and the statistics obtained for the skyrmion is apparently
not in accordance with the spin-statistics theorem. However, it must
be noted that the skyrmions in quantum Hall ferromagnets move against
a non-trivial spin background. As a result, measurement of skyrmion
spin, as a response to the external magnetic field, always have a
contribution from the background. This renders the definition of the
spin quantum number for the skyrmions somewhat arbitrary. This
situation is to be contrasted with the skyrmion in antiferromagnetic
systems like ${\rm He}^3$-A films, where the ground state has no net
spin and the skyrmion has a well defined spin quantum number in
accordance with the spin-statistics theorem.

\section {Fractional Filling }
\label{secfra}

The results of the previous section can be generalized to the case of
fractional fillings. This is most easily done within the framework of
Chern-Simons (CS) theory for fractional quantum Hall states
\cite{Jain,Lopez}. Within this framework, the electrons, apart from
their mutual interaction, are also coupled to a CS gauge field and the
coupling constant of the CS field is chosen so that an even number of
flux quanta gets attached to each electron. The resultant composite
Fermions then fill up an integer number of Landau level at the mean
field level. Thus fractional quantum Hall effect for the electrons
turns out to be integer quantum Hall effect for the composite
Fermions. In this work, we shall consider only those filling fractions
where the unperturbed ground state is known to be fully
spin-polarized. Such filling fractions are given by $\nu = 1/(1+2s)$
\cite{Gimm} where $s$ is an integer, $2s$ flux quanta are attached to
each electron and the resultant composite Fermions fill up one Landau
level. In what follows, we neglect the correlation effects between the
composite Fermions.

We shall now generalize the results of the earlier section. The procedure
is mostly similar to that of the earlier sections, and we shall only 
point out the essential differences. We start with the mean field effective
action (\ref{mean field}) which now reads
                    \begin{eqnarray}
S[\psi^{\dagger},\psi, A_{\mu}^{\rm cs}] 
&=& \int d^2r\, dt\, \psi^{\dagger}_a ({\bf r},t) 
\Bigg ( [i \partial_0 - A_{0}^{\rm cs}({\bf r},t)+\epsilon_F ] I 
- \frac {\left[ {\hat {\bf p}} - {\bf A}^{\rm eff}({\bf r}) 
- {\bf A}^{\rm cs} \right]^2} {2 m} I  \nonumber\\
&& + \gamma_0 {\bf d} ({\bf r},t) 
\cdot {\bbox \sigma} \Bigg )_{ab} 
\psi_b ({\bf r},t)  + \frac{1}{16\pi s} \int d^2r\, dt\,
\epsilon^{\mu \nu \lambda} A_{\mu}^{\rm cs}({\bf r},t)
F_{\nu \lambda}^{\rm cs}({\bf r},t),
                     \end{eqnarray}
\noindent
where $\psi$ are two component composite Fermion spinor fields. Here
${\bf A}^{\rm cs}$ means fluctuations of the CS fields from the mean
value $\langle {\bf A}^{\rm cs} \rangle $. The effective vector
potential ${\bf A}^{\rm eff}$ is then given by ${\bf A}^{\rm eff} = e
{\bf A} + \langle {\bf A}^{\rm cs} \rangle $ and the last term in
the action is the usual CS term with $F_{\mu \nu }^{\rm cs} =
\partial_{\mu} A_{\nu}^{\rm cs} - \partial_{\nu} A_{\mu}^{\rm
cs}$. The effective magnetic field experienced by the composite
Fermions are ${\bf H}_{\rm eff}= {\bf H_0}/(1+2s)$.

We now introduce SU(2) rotation matrix $U({\bf r},t)$ and the
parameters $\phi_x$ and $\phi_y$ exactly in the same fashion. After
these transformations the action $S[\chi^{\dagger},\chi,
Q_{\mu},A_{\mu}^{\rm cs}] $ can be written as
                       \begin{eqnarray}
S &=& S_0 + S_1 + S_{\rm cs}, \nonumber\\
S_0 &=& \int d^2r\, \frac{d\omega}{2 \pi}\,
\chi ^{\dagger}_a ({\bf r},\omega) \left (
 (\omega +\epsilon_F) I   
- \frac{[{\hat {\bf p}} - e {\bf A}^{\rm eff} ({\bf r}) 
 + {\bbox{\alpha}}]^2}{2 m} I 
+ \gamma_0 \sigma_z \right )_{ab} 
\chi_b ({\bf r},\omega), \nonumber\\ 
S_1 &=& - \int d^2 r\,  \frac{d \omega dp_0}{(2 \pi)^2}\,
 \chi^{\dagger}_a ({\bf r},\omega +p_0) \
 \left ( \frac{1}{2} 
\left[ ( Q_{\mu}^{\rm int}+ A_{\mu}^{\rm cs}I ),
 \frac{\partial G_0^{-1}}
{\partial \alpha_{\mu}} \right]_+ \right)_{ab} 
 \chi_b ({\bf r},\omega), \nonumber\\
S_{\rm cs} &=& \frac{1}{16\pi s} \int d^2{\bf r}\, dt\,
\epsilon^{\mu \nu \lambda} A_{\mu}^{\rm cs}({\bf r},t)
F_{\nu \lambda}^{\rm cs}({\bf r},t).
                        \end{eqnarray}
\noindent 
So, we find that the effect of fluctuations of the CS gauge fields is
to modify the basic interaction vertex $S_1$ by changing
$Q_{\mu}^{\rm int}$ to $Q_{\mu}^{\rm eff} = Q_{\mu}^{\rm int} +
A_{\mu}^{\rm cs}I$. To obtain the effective action for the skyrmion,
we therefore integrate out the Fermion fields and expand in powers of
$Q_{\mu}^{\rm eff}$. The relevant terms contributing to the Hopf term
are again given by the same diagrams in Fig.\ \ref{Graph1} with ${\bf Q}^{\rm
int}$ replaced by ${\bf Q}^{\rm eff}$. The crucial point which makes
the calculation of these relevant terms simple is that the unperturbed
Green functions are diagonal in the spin space. As a result, any term
in $S_2$ which has the form $A_{\mu}^{\rm cs} \partial_{\nu}
\Omega_{\lambda}^{1(2)\,{\rm int}}$ vanishes due to the trace
operation in spin indices. Similarly, it can be shown that all terms
involving product 
of $A_{\mu}^{\rm cs}$ and $\Omega^{\rm int}$ in $S_3$ also
vanishes. So, the cancelation of the terms in $S_2$ and $S_3$ follows
exactly in the same way as in Sec.\ref{secexp}, and finally we are
left with the effective action
                          \begin{eqnarray}
S_{\rm eff} [B_{\mu},A_{\mu}^{\rm cs}] &=& 
\frac{N_2}{32 \pi} \int d^2r\, dt \,
\epsilon^{\mu \nu \lambda} B_{\mu} F_{\nu \lambda} 
 + \frac{N_3}{8 \pi} \int d^2r\, dt \,
\epsilon^{\mu \nu \lambda} A_{\mu}^{\rm cs} F_{\nu \lambda}  
\nonumber\\
&& + (\frac{N_2}{8\pi} + \frac{1}{16 \pi s})
\int d^2r \, dt\, \epsilon^{\mu \nu \lambda} 
A_{\mu}^{\rm cs} F_{\nu \lambda}^{\rm cs}, 
                           \end{eqnarray}
\noindent
where the coefficients $N_2$ and $N_3$ are given by
                      \begin{eqnarray}
N_2 &=& 2 \pi {\rm Tr }
\left(\frac{\partial G_0^{-1}}{\partial \omega} 
G_0 \frac{\partial G_0^{-1}}{\partial \phi_x} G_0
\frac{\partial G_0^{-1}}{\partial \phi_y} 
G_0 \right),\nonumber\\
N_3 &=& 2 \pi {\rm Tr }
\left(\sigma_3 \frac{\partial G_0^{-1}}{\partial \omega} 
G_0 \frac{\partial G_0^{-1}}{\partial \phi_x} G_0
\frac{\partial G_0^{-1}}{\partial \phi_y} G_0 \right).
                       \end{eqnarray}

Since the unperturbed ground state for the composite Fermions is
completely spin polarized, it can be easily seen that $N_2 =
N_3$. Also since the composite Fermions fill up exactly one effective
Landau level for the considered filling fractions, the arguments of
Sec.\ref{seceva} can again be carried through to get $N_2 = N_3 = 1$.
Using these results and integrating out the CS gauge fields
\cite{Lopez}, we finally get
                       \begin{eqnarray}
S_{\rm Hopf} &=& \frac{\nu}{32 \pi} \int d^2r\, dt\, 
\epsilon^{\mu \nu \lambda} B_{\mu} F_{\nu \lambda},  
                        \end{eqnarray}
where $\nu = 1/(1+2s)$ is the filling fraction.

So, all the results of the previous sections can be generalized by
replacing the value of $N$ by $\nu$. The skyrmions are anyons 
carrying a charge $e\nu Q_{\rm top}$.  This coincides with the
results obtained by Baez ${\it et\,al.}$ \cite{Baez} using a
phenomenological approach.

There are a couple of points that are worth mentioning here. First,
the same results may be obtained using the formalism of spin-allowed
CS theory \cite{Gimm}. The calculations follow exactly the same lines
for the filing fractions discussed here, and are therefore not
repeated. Second, although the theoretical derivations are exactly the
same as in the case of integer filling, fractional fillings are
achieved at much stronger external fields which means that
experimental observation of skyrmions for fractional fillings will
require samples with very low Lande g factor. Recently, however,
Leadly $\it{et\,al.}$ \cite{Leadly} have observed evidence of spin
depolarization near $\nu = 1/3$ by reducing the Lande g factor of the
sample by applying external pressure.

\section { Finite Temperature}
\label{secfin}

In this section, we derive the effective action for the skyrmions at
finite temperature at $\nu = 1$.  We shall also assume that disorder
is negligible ${\it i.e.}$ $\omega_c \tau \gg 1$ and the temperature
$T = 1/\beta$ is small enough so that one can neglect
phonons. Throughout this section, we shall use the Matsubara
formalism. We use natural units $\hbar = c = k_B = 1$ and all other
conventions listed in Sec. \ref{seceva} remain the same except for the
replacement $ t \rightarrow -i\tau$.

The action for the system can be written using the same model of local
interaction as in Sec \ref{secexp} (\ref{mean field},\ref{mfh}), but
with $i \int dt$ replaced by $-\int 
\limits_{0}^{\beta} d \tau $. Following steps identical to those in
Sec.\ref{secexp}, we then integrate out the Fermionic fields to obtain
the effective action given by
                     \begin{eqnarray}
S_{\rm eff} &=& S_2 + S_3, \nonumber\\
S_2 &=& \frac{1}{2} \int \frac{d^2 p }{(2 \pi)^2}\,
\frac{1}{\beta}\sum_{ip_n} 
\left [\Omega_{\mu}^{1\,{\rm int}} ({\bf p},ip_n) 
 P^{\mu \nu} _{1} ({\bf p},ip_n) 
\Omega_{\nu}^{1\,{\rm int}}(-{\bf p},-ip_n) 
\right. \nonumber\\
&& \left. + \Omega_{\mu}^{2\,{\rm int}} ({\bf p},ip_n) 
 P^{\mu \nu} _{2} ({\bf p},ip_n) 
\Omega_{\nu}^{2\,{\rm int}}(-{\bf p},-ip_n) 
+\Omega_{\mu}^{3\,{\rm int}} ({\bf p},ip_n) 
 P^{\mu \nu} _{3} ({\bf p},ip_n) 
\Omega_{\nu}^{3\,{\rm int}}(-{\bf p},-ip_n) 
\right ] \nonumber\\
S_3 &=& \int \frac{d^2 p \,d^2 q \,d^2 l }{(2\pi)^6} \,
\frac{1}{\beta^2} \sum_{ip_n, iq_n}
T^{\mu \nu \lambda}_{123}({\bf p},{\bf q}, ip_n, iq_n)
 \Omega_{\mu}^{1\,{\rm int}} ({\bf p},ip_n) 
\Omega_{\nu}^{2\,{\rm int}} ({\bf q},iq_n) \nonumber\\
&& \times \Omega_{\lambda}^{3\,{\rm int}} 
\left(-{\bf p}-{\bf q}),- ip_n - iq_n \right),
\label{finite temp}
                        \end{eqnarray}
where $p_n$ and $q_n$ are the Matsubara frequencies and $P^{\mu \nu}
_{i}$ and $T^{\mu \nu \lambda}_{123}$ can be expressed in terms of the
unperturbed Green functions $G_0$ as
                      \begin{eqnarray}
P^{\mu \nu} _{i} ({\bf p},ip_n) &=& \frac{-1}{4 L_x L_y}
{\rm Tr}\left ( \int d^2 r_1 d^2 r_2  \frac{1}{\beta} \sum_{i \omega_n} 
\,\frac{1}{2} \left[ \sigma_i e^{i {\bf p}\cdot {\bf r_1} },
\left(\frac{\partial G_0^{-1}}{\partial \alpha_{\mu}}\right)
_{{\bf r}_1} \right]_+ 
G_0 ({\bf r_1},{\bf r_2},i\omega_n) \right. \nonumber\\
&&\left. \times 
\frac{1}{2} \left [ \sigma_i e^{-i {\bf p}\cdot {\bf r_2}}, 
\left(\frac{\partial G_0^{-1}}{\partial \alpha_{\nu}}\right)
_{{\bf r}_2} \right]_+ 
G_0 ({\bf r_2},{\bf r_1},i\omega_n + ip_n) \right ), \nonumber\\
T^{\mu \nu \lambda}_{123}
({\bf p},{\bf q}, ip_n, iq_n)&=& \frac{i}{4 L_x L_y}
{\rm Tr} \left( \int d^2 r_1 d^2 r_2 d^2 r_3 
\frac{1}{\beta} \sum_{i\omega_n} 
\frac{1}{2} \left [ \sigma_1 e^{i {\bf p}\cdot {\bf r_1}},
\left(\frac{\partial G_0^{-1}}
{\partial \alpha_{\mu}}\right)_{{\bf r}_1} \right]_+ 
 \right. \nonumber\\
&& \left.  \times G_0 ({\bf r_1},{\bf r_2},i\omega_n ) 
\frac{1}{2} \left 
[ \sigma_2 e^{i {\bf q}\cdot {\bf r_2}},
\left(\frac{\partial G_0^{-1}}
{\partial \alpha_{\nu}}\right)_{{\bf r}_2} \right]_+
G_0 ({\bf r_2},{\bf r_3},i\omega_n +iq_n)
\right. \nonumber\\
&& \left.  \times
\frac{1}{2} \left [ \sigma_3 e^{-i ({\bf p}+{\bf q})\cdot {\bf r_3}},
\left(\frac{\partial G_0^{-1}}
{\partial \alpha_{\lambda}}\right)_{{\bf r}_3} \right]_+ 
G_0 \left({\bf r_3},{\bf r_1},i\omega_n -ip_n \right) \right).
                    \end {eqnarray}

First, we consider the  terms in the effective action which
contain $T^{\mu \nu \lambda}_{123}$, $ P^{\mu \nu} _{1}$ and $P^{\mu
\nu} _{2}$. It can be easily shown that each of these terms are
analytic functions of ${\bf p}$ and $p_0$. The reason for this
analyticity is simple. Each of these functions contain a product of
two or three Green functions in combination of $G_0^{\pm}G_0^{\mp}$ or
$G_0^{\pm}G_0^{\pm}G_0^{\mp}$. As a result, the energy eigenvalues in
the denominator of these Green functions are different even when their
Landau level indices are the same and this restores the analyticity of
these functions. Since these functions are analytic, we can carry out
a derivative expansion similar to the zero temperature case. This
leads to cancelation of the first two terms of $S_2$ and $S_3$ up to
the required order.

The procedure for obtaining the effective action has been, till now,
similar to that in the zero temperature case. However, at this stage,
we come across an important difference. At finite temperature, $P^{\mu
\nu} _{3}$ turns out to be a non-analytic function of frequency and
momenta, so that we can not carry out a derivative expansion analogous
to the zero temperature case to obtain the Hopf term. This
non-analyticity of the polarization bubble is well known in finite
temperature field theory and many-body theory
\cite{Das,Aitchison,Kao}. To see exactly where this non-analyticity
appear, we compute $P^{\mu \nu} _{3}({\bf p},p_0)$. The method of this
computation is sketched in App.C. The result is
                       \begin{eqnarray}
P^{00}_3({\bf p},p_0) &=& |{\bf p}|^2 \Pi_0({\bf p},p_0),\nonumber\\
P^{i0}_3({\bf p},p_0) &=& p_i p_0 \Pi_0({\bf p},p_0) 
+ i \epsilon^{i0j} p_j \Pi_1({\bf p},p_0), \nonumber\\
P^{ij}_3({\bf p},p_0) &=& p_0^2 \Pi_0({\bf p},p_0) 
+  i \epsilon^{ij0} p_0 \Pi_1({\bf p},p_0) \nonumber\\
&& +(|{\bf p}|^2 \delta_{ij} - p_i p_j) \Pi_2({\bf p},p_0),
\label{pmunu}
\end{eqnarray}
where the functions $\Pi_0$, $\Pi_1$ and $\Pi_2$ are given by
\begin{eqnarray}
\Pi_0({\bf p},p_0) &=& \frac{-l_B^2}{4} 
\sum_{n = 0}^{\infty} \sum_{n' = 0}^{n}
\sum_{\alpha = \pm}
C_{nn'}^{\alpha}(p_0) \frac{n'!}{ n!}  e^{\frac{-|{\bf p}|^2 l_B^2}{2}}
\left(\frac{|{\bf p}|^2 l_B^2}{2}\right)^{n-n'-1} 
L_{n'}^{n-n'}\left(\frac{|{\bf p}|^2 l_B^2}{2}\right), \nonumber\\
\Pi_1({\bf p},p_0) &=& \frac{-l_B \omega_c^{\frac{1}{2}}
}{ 8\sqrt{ m}} 
\sum_{n = 0}^{\infty} \sum_{n' = 0}^{n} \sum_{\alpha=\pm}
C_{nn'}^{\alpha}(p_0)
\frac{n'!}{ n!}  e^{\frac{-|{\bf p}|^2 l_B^2}{2}}
\left(\frac{|{\bf p}|^2 l_B^2}{2}\right)^{n-n'-1} 
L_{n'}^{n-n'}\left(\frac{|{\bf p}|^2 l_B^2}{2}\right) \nonumber\\
&& \times \left [
\frac{|{\bf p}|^2 l_B^2}{2}
\left\{ L_{n'}^{n-n'}\left(\frac{|{\bf p}|^2 l_B^2}{2}\right) 
+ 2 L_{n'-1}^{n-n'+1}\left(\frac{|{\bf p}|^2 l_B^2}{2}\right)
\right \} - (n-n') L_{n'}^{n-n'}\left(\frac{|{\bf p}|^2 l_B^2}{2}
\right) \right ], \nonumber\\
\Pi_2({\bf p},p_0) &=& \frac{-l_B^2}{16 m} 
\sum_{n = 0}^{\infty} \sum_{n' = 0}^{n}
\sum_{\alpha = \pm}
C_{nn'}^{\alpha}(p_0)
\frac{n'!}{ n!} e^{\frac{-|{\bf p}|^2 l_B^2}{2}}
\left(\frac{|{\bf p}|^2 l_B^2}{2}\right)^{n-n'-1} \nonumber\\
&& \times
\left\{L_{n'}^{n-n'}\left(\frac{|{\bf p}|^2 l_B^2}{2}\right) +
2 L_{n'-1}^{n-n'+1}\left(\frac{|{\bf p}|^2 l_B^2}{2}\right) \right \}
\nonumber\\
&& \times \left [\frac{|{\bf p}|^2 l_B^2}{2}
\left \{ L_{n'}^{n-n'}\left(\frac{|{\bf p}|^2 l_B^2}{2}\right) 
+ 2 L_{n'-1}^{n-n'+1}\left(\frac{|{\bf p}|^2 l_B^2}{2}\right)
\right \}  - (n-n') L_{n'}^{n-n'}\left(\frac{|{\bf p}|^2 l_B^2}{2}
\right) \right ], \nonumber\\
\label{pifunction}
\end{eqnarray}
where $n$ and $n'$ are Landau level indices, $f$ is the Fermi
distribution function, $f_n^{\alpha}$ is shorthand notation for
$f(\epsilon_n^{\alpha})$, $\epsilon_n^{\alpha}$ are the energy
eigenvalues given by $ \epsilon_n^{\alpha} = (n+\frac{1}{2})\omega_c +
\alpha \gamma_0$, $L_{n'}^{n-n'}$ is the associated Laugurre
polynomial and $C_{nn'}^{\alpha}$ is given by 
                       \begin{eqnarray}
C_{nn'}^{\alpha}(p_0) &=& \left(\frac{(f_n^{\alpha}-f_{n'}^{\alpha})
\omega_c(n-n')}{p_0^2 -\omega_c^2(n-n')^2} \right) \,\,\,\,\,\,\,
{\rm if} \,\,\,p_0 \ne 0 ,\nonumber\\ 
&=& -\delta_{nn'}
\left(\frac{\partial f(z)}{\partial z}\right)_{z =
\epsilon_n^{\alpha}} \,\,\,\,\,\,\,\,\,
{\rm if} \,\,\,p_0 = 0.
\label{freqsum}
                        \end{eqnarray}
 
From Eq.\ (\ref{pifunction}), it can be shown that the functions
$\Pi_1$ and $\Pi_2$ are non-analytic at the origin. The value of these
functions in the limit ${\bf p} \rightarrow 0$ depends on whether
$p_0$ is set to zero before or after the limit is taken. This non
-analytic nature arises from the contribution of the $ n = n'$ term in
the expression of $C_{nn'}^{\alpha}(p_0)$. Further, $\Pi_0$, $\Pi_1$
and $\Pi_2$ do not either diverge or approach zero as ${\bf p}$ and
$p_0 \rightarrow 0$ irrespective of how the limit ${\bf p}\rightarrow
0, p_0 \rightarrow 0$ is taken. So, it can be inferred from
Eqs.\ (\ref{finite temp}), (\ref{pmunu}) and (\ref{pifunction}) that in
the long-wavelength low-frequency limit, the relevant leading order
term in the effective action is given by
                           \begin{eqnarray} 
S_{\rm eff} &=& \frac{1}{2} \int \frac{d^2p \,d
p_0}{(2\pi)^3} \,\epsilon^{\mu \nu \lambda} B_{\mu}^{{\rm int}} ({\bf
p},p_0) i p_{\lambda} \Pi_1({\bf p},p_0)
B_{\nu}^{{\rm int}}(-{\bf p},-p_0),
\label{effac}
                          \end{eqnarray}
where we have expressed $\Omega_{\mu}^{3\,{\rm int}}$ fields in terms
of the gauge fields $B_{\mu}^{{\rm int}}$ using Eq.\
(\ref{rotation}). If we could have taken an unambiguous ${\bf
p}\to 0, p_0 \to 0$ limit for $\Pi_1$, Eq.\ (\ref{effac}) would reduce
to a local effective action in position space similar Eq.\
(\ref{Hopfterm}). However, such a limit does not exist and in general
it is not possible to obtain a simple effective action in position
space. Nevertheless, as shown above, it is still possible to obtain a rather
simple effective action in momentum space.

Although, it is not possible to take an unambiguous zero frequency zero
momentum limit of the effective action (\ref{effac}), one can
still infer the fate of the prefactor $N$ at finite temperature from
physical considerations. To elucidate this point, let us consider the
Hall conductivity (which is the Chern number) in these systems at
finite temperature and in absence of disorder. If we attempt to
compute the term in the effective action giving rise to the Hall
conductivity, we find that we obtain an identical non-analytic
polarization tensor and the effective action is again given by
                  \begin{eqnarray} 
S_{\rm eff}^{\rm Hall} &=& \frac{e^2}{2} \int
\frac{d^2p \,d p_0}{(2\pi)^3} \,\epsilon^{\mu \nu \lambda} A_{\mu}^{\rm
em} ({\bf p},p_0) i p_{\lambda} \Pi_1({\bf
p},p_0) A_{\nu}^{\rm em}(-{\bf p},-p_0).
\label{Hallac}
                  \end{eqnarray} 
To obtain the d.c Hall conductivity as a response of the system to an
external field from this effective action, we must now specify the dc
limit. There are two different limits possible. In the dynamic limit,
we measure the response of the system to a spatially homogeneous but
very slowly time varying external field. In this case, to compute the
response, we first set ${\bf p}=0$ and then take the limit $p_0
\rightarrow 0$. On the other hand, in the static limit, we measure the
response of the system with respect to a static but weakly spatially
inhomogeneous field and consequently first set $p_0 = 0$ and then take
the ${\bf p} \rightarrow 0$ limit. It is well known that at finite
temperature the conductivities in the static and the dynamic limits
are different \cite{Sakhi,Zhang,Ingraham}.

Similarly, the prefactor of the Hopf term, which contributes in part
to the spin of the skyrmion, can be looked upon as the response of the
system to external magnetic field \cite{Volovik1}. In particular, part
of the z component of the spin ($J_z^{\rm Hopf}$) can be obtained from
the Hopf term as a response to an external magnetic field in the z
direction. It can be easily seen, that the Hopf term in the effective
action of the system with the external magnetic field ( $H^{\rm
ext}_z)$ is given by Eq.\ (\ref{effac}) with $B_{\mu}^{{\rm int}}$
replaced by $B_{\mu}^{{\rm total}} = B_{\mu}^{{\rm int}}+ H^{\rm
ext}_z $. The contribution from the Hopf term to the z component of
the spin of the skyrmion can then be computed from this action as a
response of the system to the external magnetic field $H^{\rm
ext}_z$. To compute this, we must again specify the dc limit. In the
dynamic limit, relevant for spatially homogeneous magnetic field with
weak temporal variation, a simple calculation leads to $J_z^{\rm 
Hopf} = N_{\rm dynamic}/2$, where
                      \begin{eqnarray}
N_{\rm dynamic} &=& 16 \pi \lim_{p_0 \to 0} \Pi_1(0,p_0)
= N = 1. 
                       \end{eqnarray}
The prefactor of the Hopf term remain the same as the zero temperature
value in the dynamic limit.  In the static limit, appropriate for
static but weakly spatially varying magnetic field, we get $J_z^{\rm
Hopf} = N_{\rm static}/2 $, where
                     \begin{eqnarray}
N_{\rm static} &=& 16 \pi \lim_{{\bf p} \to 0} 
\Pi_1({\bf p},0)
= 1 - \beta \omega_c
\sum_{n =0}^{\infty} \sum_{\alpha = \pm} (n+\frac{1}{2}) 
f_n^{\alpha}(1-f_n^{\alpha}).
\label{staticN}
                     \end{eqnarray}
The prefactor of the Hopf term becomes a function of temperature in
the static limit. The temperature dependence of $N_{\rm static}$ is
shown in Fig.\ \ref{Graph2}. As expected, $N_{\rm static} = N_{\rm
dynamic}$ at $T=0$ and reduces to zero at high temperature
($T/\omega_c\gg 1$). The temperature dependence of $N_{\rm dynamic}$
and $N_{\rm static}$ is similar to that of the dynamic and static Hall
conductivities in these systems \cite{Sakhi,Zhang,Ingraham}.

\section  {Conclusion}
\label{seccon}

The present work generalizes the derivation of the Hopf term in
earlier works \cite{Volovik1,Yakovenko1,Yakovenko2} to quantum Hall
systems. We find that although $k_x$ and $k_y$ are no longer good
quantum numbers simultaneously, it is possible to replace the
integration over these wave vectors by averaging over phases of
the boundary conditions. This procedure is well known and is very
useful in determining the quantized Hall conductivity which is
essentially given by the same topological invariant
\cite{Niu,Kohomoto}. The prefactor of the Hopf term ($N$) is found to
be the filling factor ($\nu$).

The value of $N$ obtained agrees to the previous result
\cite{Apel,Ray,Iordanskii} for the integer filling case. The method
used here is, however, quite different and more powerful. It does not
need either the assumption of lowest Landau level projection
\cite{Apel,Ray} or laborious term by term evaluation of the effective
action\cite{Iordanskii}.  Furthermore it provides a simple physical
argument in terms of Berry's phase as to why the prefactor the Hopf
term has to be an integer. It is pointed out that this is a general
result, and not a consequence of the simplicity of the model
assumed. Our method also does not require parameterization of the
SU(2) rotation matrix $U$ in terms of the Euler angles and therefore
avoids any ambiguity which may arise from such a procedure
\cite{Volovik2,Volovik3}. The generalization of this method to the
case of fractional filling factors $\nu = 1/(2s+1)$ is
straightforward. The results 
obtained are in agreement with the results of Baez ${\it et\,al.}$
\cite{Baez} obtained using phenomenological approach.

The expressions for the spin and charge densities of the skyrmion are
obtained from the effective action. The contribution to the spin
density comes from both the Berry term and the Hopf term in the
effective action. In the case of the long-wavelength skyrmions, the
contribution due to the former dominates.  The total spin of the
skyrmion, therefore, depends on the system details. The charge and the
statistics of the skyrmion, on the other hand, is completely
determined by the prefactor of the Hopf term. The charge for a
skyrmion with $Q_{\rm top}=1$ is shown to be $\nu e$ and the
statistical phase is found to be $\nu \pi$.  Consequently, the
skyrmions are Fermions (anyons) for odd integer (fractional) fillings.

We also obtain an effective action for the skyrmion at finite
temperature. It is shown that it is not, in general, possible to
obtain a local Hopf term in position space at finite temperature
because of the non-analyticity of $P^{\mu \nu} _{3}({\bf
p},p_0)$. However, it is possible to derive a rather simple effective
action for the system in the momentum space. One can then choose
static or dynamic limits based on physical reasoning and find the
value of the Hopf invariant in these limits. In the dynamic limit, the
prefactor of the Hopf term is independent of temperature and has the
same value as at zero temperature while in the static limit it becomes
a function of temperature.

To conclude, we derive the prefactor of the Hopf term in the effective
action of skyrmions in quantum Hall systems for both integer and
fractional fillings. This prefactor ($N$) is found to be $\nu$. This
suggests that the skyrmion with $Q_{\rm top}=1$ have charge $\nu e$
and statistical phase $\nu \pi$. On the other hand, the dominant
contribution to the spin of the skyrmion comes from the Berry term in
the effective action. As a result, the spin of the skyrmion depends on
system parameters and increases with the size of the skyrmion. We also
discuss the fate of the Hopf term at finite temperature.  At finite
temperature, it is in general not possible to obtain a local Hopf term
in position space. However, it is possible to obtain an effective
action in the momentum space and to obtain the value of the prefactor
of the Hopf term from this action in the static and the dynamic limit.

The authors would like to thank G. E. Volovik and M. Stone for helpful
discussion.

\section{Appendix A}

In this section, we briefly outline the calculation of the relevant
term in the action from the expression of $P^{\mu \nu} _{ij} ({\bf
p},p_0)$. For the relevant term in the effective action, we
need to retain the first order term in expansion of $P^{\mu \nu} _{ij}
({\bf p},p_0)$. We start from the expression of $P^{\mu \nu}
_{ij} $ in Eq.\ (\ref{Pi tensor}) which is
             \begin{eqnarray}
P^{\mu \nu} _{ij} ({\bf p},p_0) &=& \frac{i}{4 L_x L_y}
{\rm Tr} \left( \int d^2 r_1 \,d^2 r_2  \frac{d \omega}{2 \pi} 
\frac{1}{2} \left[ \sigma_i e^{i {\bf p}\cdot {\bf r_1} },
\left(\frac{\partial G_0^{-1}}{\partial \alpha_{\mu}}\right)
_{{\bf r}_1} \right]_+ 
G_0 ({\bf r_1},{\bf r_2},\omega) \right. \nonumber\\
&&\left. \times 
\frac{1}{2} \left [ \sigma_j e^{-i {\bf p}\cdot {\bf r_2}}, 
\left(\frac{\partial G_0^{-1}}{\partial \alpha_{\nu}}\right)
_{{\bf r}_2} \right]_+ 
G_0 ({\bf r_2},{\bf r_1},\omega + p_0) \right ),
\label{pol tensor}
              \end{eqnarray}
where $G_0$ is the unperturbed Green function.   Using the Fourier
transform of the Green functions,
                  \begin{eqnarray}
G_0 ({\bf k}_1,{\bf k}_2,\omega ) &=& \int d^2 r_1\, d^2 r_2 
e^{i ({\bf k}_1 \cdot {\bf r}_1 - {\bf
k}_2 \cdot {\bf r}_2)} G_0( {\bf r}_1,{\bf r}_2,\omega),\nonumber
                   \end{eqnarray}
Eq.\ (\ref{pol tensor}) can be written in Fourier space as
                   \begin{eqnarray}
P^{\mu \nu} _{ij}({\bf p},p_0)&=& \frac{i}{4 L_x L_y} {\rm Tr} 
\int \frac{d^2 k_1 \,d^2 k_2 \,d \omega}{(2\pi)^5} 
\left (\sigma_i \left( \frac{\partial G_0^{-1}}
{\partial \alpha_{\mu}} \right)_{{\bf k}_1 +\frac{ {\bf p}}{2}}
 G_0 ( {\bf k}_1 , {\bf k}_2 , \omega  ) \right. \nonumber\\
&& \left. \times \sigma_j \left( \frac{\partial G_0^{-1}}
{\partial \alpha_{\nu}} \right)_{{\bf k}_2 + \frac{{\bf p}}{2}}
 G_0 ( {\bf k}_2 + {\bf p}, {\bf k}_1 
+ {\bf p}, \omega + p_0 ) \right).
                   \end{eqnarray}
Next, we carry out a Taylor expansion of $P^{\mu \nu}$ in powers of the
external frequency and momentum $p_{\lambda}$, where $\lambda$ can
take values (0,1,2). After some tedious but straightforward
manipulations we finally obtain
                      \begin{eqnarray}
P^{\mu \nu} _{ij}({\bf p},{\bf q},p_0) &=&  
\frac{-ip_{\lambda}}{4 L_x L_y}
\int \frac{d^2 k_1 \,d^2 k_2 \,d^2 k_3 \,d \omega}{(2\pi)^7} 
{\rm Tr} \left (\sigma_i \left( \frac{\partial G_0^{-1}}
{\partial \alpha_{\mu}} \right)_{{\bf k}_1} 
 G_0 ( {\bf k}_1 , {\bf k}_2 , \omega  ) 
\sigma_j \right. \nonumber\\
&& \left. \times \left( \frac{\partial G_0^{-1}}
{\partial \alpha_{\nu}} \right)_{{\bf k}_2}
 G_0 ( {\bf k}_2 , {\bf k}_3 ,\omega ) 
\left( \frac{\partial G_0^{-1}}
{\partial \alpha_{\lambda}} \right)_{{\bf k}_3}
 G_0 ( {\bf k}_3, {\bf k}_1,\omega )\right), 
                  \end{eqnarray}
\noindent
where  we have used the identities following from Eq.\ (\ref{geq})
              \begin{eqnarray}
\frac{\partial G_0 ( {\bf k}_1, {\bf k}_2, \omega )}
{\partial  k _1^a} +\frac{\partial G_0 ( {\bf k}_1, 
{\bf k}_2, \omega )}{\partial  k _2^a}
&=&  \int \frac{d^2 k_3}{(2\pi)^2} 
G_0 ( {\bf k}_1, {\bf k}_3,\omega )
\left( \frac{\partial G_0^{-1}}
{\partial \alpha_a} \right)_{{\bf k}_3}
 G_0 ( {\bf k}_3, {\bf k}_2,\omega ), \nonumber\\
\frac{\partial G_0 ( {\bf k}_1, {\bf k}_2, \omega )}
{\partial \omega} 
&=& - \int \frac{d^2 k_3}{(2\pi)^2}
G_0 ( {\bf k}_1, {\bf k}_3,\omega )
\left( \frac{\partial G_0^{-1}}
{\partial \alpha_0} \right)_{{\bf k}_3}
G_0 ( {\bf k}_3, {\bf k}_2,\omega ). 
              \end{eqnarray}

Substituting the expression of $P^{\mu \nu}_{ij}$ in Eq.\ (\ref{second
order}), and transforming back to real space, we obtain for $S_2$
                       \begin{eqnarray}
S_2 &=&  \frac{1}{2} b^{\mu \nu \lambda}_{ij} 
 \int d^2r\, dt \,\Omega_{\nu}^{j \,{\rm int}}({\bf r},t) 
\partial_{\lambda} 
\Omega_{\mu}^{i\,{\rm int}} ({\bf r},t), 
                        \end{eqnarray}
where the tensor $ b^{\mu \nu \lambda}_{ij}$ is given by
\begin{eqnarray}
b^{\mu \nu \lambda}_{ij} &=& \frac{1}{4 L_x L_y} {\rm Tr} 
\left ( \sigma_i \frac{\partial G_0^{-1}}
{\partial \alpha_{\mu}} G_0 
 \sigma_j \frac{\partial G_0^{-1}} 
 {\partial \alpha_{\nu}} G_0 
 \frac{\partial G_0^{-1}}
{\partial \alpha_{\lambda}} G_0 \right ). 
\end{eqnarray}

This procedure can be similarly carried out to obtain the zeroth order
term in expansion of $T^{\mu \nu \lambda}_{ijk}$ which contribute to
$S_3$. The manipulations are straightforward and directly yield Eq.\
(\ref{Triangle Graph}).

\section {Appendix B}

In this section, we sketch the derivation of some of the properties of
the tensors $b^{\mu \nu \lambda}_{ij}$ and $c^{\mu \nu
\lambda}_{ijk}$ which lead to cancelation of the terms $S_2$ and
$S_3$ in the effective action. We begin with the tensor $c^{\mu \nu
\lambda}_{ijk}$. It is easy to see from the expression of $c^{\mu \nu
\lambda}_{ijk}$ in Eq.\ (\ref{Triangle Graph}), that all components of
this tensor that have odd number of $\sigma_1$ or $\sigma_2$ matrices
vanish by trace operation. So the only possible non-zero terms in
$S_3$ are of the form
\begin{eqnarray}
S_3 &=&  \frac{1}{6} c^{\mu \nu \lambda}_{ij3} 
\int d^2 r \,dt \,\Omega_{\mu}^{i\,{\rm int}} ({\bf r},t) 
\Omega_{\nu}^{j\,{\rm int}} ({\bf r},t) 
\Omega_{\lambda}^{3\,{\rm int}} ({\bf r},t), 
\label{symac}
\end{eqnarray}
where the indices $i$ and $j$ are such that $\sigma_i \sigma_j = I$ or
$\pm i \sigma_3$. We first consider the case where $i$ = $j$. Let us
perform a gauge transformation on $S_3$, which corresponds, as we have
seen in Sec.(\ref{secspi}), to a rotation about the local ${\bf d}$
axis. Under such a transformation, $\Omega_{\lambda}^{3\,{\rm int}}
\to \Omega_{\lambda}^{3\,{\rm int}}+ \partial_{\lambda} \Lambda$ and
the action $S_3$ picks up an extra term
\begin{eqnarray}
\delta S_3 &=& \frac{1}{6} c^{\mu \nu \lambda}_{ii3} 
\int d^2 r \,dt \,\Omega_{\mu}^{i\,{\rm int}} ({\bf r},t) 
\Omega_{\nu}^{i\,{\rm int}} ({\bf r},t) 
\partial_{\lambda} \Lambda ({\bf r},t).
\end{eqnarray}
It therefore follows that for the action to be gauge-invariant,
$\delta S_3$ must vanish for all $i$ and consequently we must have
$c^{\mu \nu \lambda}_{ii3} = -c^{\nu \mu \lambda}_{ii3}$. The property
$c^{\mu \nu \lambda}_{ii3} = -c^{\nu \mu \lambda}_{ii3} $ can also be
checked by explicit evaluation of the coefficient $c^{\mu \nu
\lambda}_{ii3}$. As a result, the terms in $S_3$ with identical spin
indices vanish. This can be seen by interchanging the indices $\mu$
and $\nu$ in Eq.\ (\ref{symac}) for terms with $i = j$ and $k = 3$. So
the only non-vanishing terms in $S_3$ are those with all different
spin indices. Furthermore, since $c^{\mu \nu \lambda}_{123} = -b^{\mu
\nu \lambda}_{1}$ (\ref{corela}), $S_3$ can be expressed as
\begin{eqnarray}
S_3 &=& - \frac{1}{2} b^{\mu \nu \lambda}_{1} \int d^2r\, dt\,
\Omega_{\lambda}^{3\,{\rm int}}\left( \Omega_{\mu}^{1\,{\rm
int}}\Omega_{\nu}^{2\,{\rm int}} - \Omega_{\mu}^{2\,{\rm int}}
\Omega_{\nu}^{1\,{\rm int}}\right).
\label{ss3}
\end{eqnarray}

Next, we demonstrate the cancelation of the terms in $S_2$ and
$S_3$. We begin with the first term of $S_2$ in
Eq.\ (\ref{action2}). Using Eq.\ (\ref{gauge}), this term can be expressed
as 
\begin{eqnarray}
S_2^1 &=& \frac{1}{2} b^{\mu \nu \lambda}_{1} 
\int d^2r\, dt\, \left[ \Omega_{\nu}^{1\,{\rm int}} \left(
\partial_{\mu} \Omega_{\lambda}^{1\,{\rm int}}
-\Omega_{\lambda}^{2\,{\rm int}}\Omega_{\mu}^{3\,{\rm int}} 
+\Omega_{\lambda}^{3\,{\rm int}}\Omega_{\mu}^{2\,{\rm int}}
\right)\right] \nonumber\\
&=& \frac{1}{2} b^{\mu \nu \lambda}_{1} 
\int d^2r\, dt\, \left[ - \Omega_{\lambda}^{1\,{\rm int}}
\left(\partial_{\mu} \Omega_{\nu}^{1\,{\rm int}} \right)
-\Omega_{\nu}^{1\,{\rm int}}
\Omega_{\lambda}^{2\,{\rm int}}\Omega_{\mu}^{3\,{\rm int}} 
+\Omega_{\nu}^{1\,{\rm int}}
\Omega_{\lambda}^{3\,{\rm int}}\Omega_{\mu}^{2\,{\rm int}}
\right], 
\label{firstterm}
\end{eqnarray}
where in the last step we have integrated the first term in $S_2^1$ by
parts. Using Eq.\ (\ref{gauge}), it can be seen after some
algebra that Eq.\ (\ref{firstterm}) can be expressed as
\begin{eqnarray}
S_2^1 &=& \frac{1}{2}\int d^2r\, dt\, {\Bigg \{} b^{\mu \nu \lambda}_{1} 
\left[ \Omega_{\mu}^{3\,{\rm
int}}\Omega_{\nu}^{2\,{\rm int}}\Omega_{\lambda}^{1\,{\rm int}} 
+ \Omega_{\mu}^{3\,{\rm int}}\Omega_{\nu}^{1\,{\rm
int}}\Omega_{\lambda}^{2\,{\rm int}} 
- \Omega_{\mu}^{2\,{\rm int}}\Omega_{\nu}^{1\,{\rm
int}}\Omega_{\lambda}^{3\,{\rm int}} \right] \nonumber\\
&& - ( b^{\mu \nu \lambda}_{1}+ b^{\nu \mu \lambda}_{1} )
\left[ \Omega_{\lambda}^{1\,{\rm int}}
\left(\partial_{\nu} \Omega_{\mu}^{1\,{\rm int}} \right) +
\frac{1}{2}\Omega_{\mu}^{3\,{\rm int}}\Omega_{\nu}^{2\,{\rm
int}}\Omega_{\lambda}^{1\,{\rm int}} \right] {\Bigg \}}.
\label{ss21}
\end{eqnarray}
An identical calculation for the second term in Eq.\ (\ref{action2})
yields
\begin{eqnarray}
S_2^2 &=& \frac{1}{2}\int d^2r\, dt\, {\Bigg \{} b^{\mu \nu
\lambda}_{1} \left[ \Omega_{\mu}^{1\,{\rm
int}}\Omega_{\nu}^{3\,{\rm int}}\Omega_{\lambda}^{2\,{\rm int}} 
+ \Omega_{\mu}^{1\,{\rm int}}\Omega_{\nu}^{2\,{\rm
int}}\Omega_{\lambda}^{3\,{\rm int}} 
- \Omega_{\mu}^{3\,{\rm int}}\Omega_{\nu}^{2\,{\rm
int}}\Omega_{\lambda}^{1\,{\rm int}}\right] \nonumber\\
&& - ( b^{\mu \nu \lambda}_{1}+ b^{\nu \mu \lambda}_{1} )
\left[ \Omega_{\lambda}^{2\,{\rm int}}
\left(\partial_{\nu} \Omega_{\mu}^{2\,{\rm int}} \right) +
\frac{1}{2}\Omega_{\mu}^{1\,{\rm int}}\Omega_{\nu}^{3\,{\rm
int}}\Omega_{\lambda}^{2\,{\rm int}} \right] {\Bigg \}}.
\label{ss22}
\end{eqnarray}
Using Eqs.\ (\ref{ss21}), (\ref{ss22}) and (\ref{ss3}), after some
straightforward algebra, one obtains
\begin{eqnarray}
S_2^1 + S_2^2 + S_3 &=& -\frac{1}{2}
( b^{\mu \nu \lambda}_{1}+ b^{\nu \mu \lambda}_{1} ) \int d^2r\, dt\, 
\Big[ \Omega_{\lambda}^{1\,{\rm int}} 
\left(\partial_{\nu} \Omega_{\mu}^{1\,{\rm int}} \right) 
+ \Omega_{\lambda}^{2\,{\rm int}}
\left(\partial_{\nu} \Omega_{\mu}^{2\,{\rm int}} \right) \nonumber\\-
&& \frac{1}{2} \left( \Omega_{\mu}^{3\,{\rm int}}\Omega_{\nu}^{2\,{\rm
int}}\Omega_{\lambda}^{1\,{\rm int}}
+ \Omega_{\mu}^{1\,{\rm int}}\Omega_{\nu}^{3\,{\rm
int}}\Omega_{\lambda}^{2\,{\rm int}} - \Omega_{\mu}^{3\,{\rm
int}}\Omega_{\nu}^{1\,{\rm int}}\Omega_{\lambda}^{2\,{\rm int}}\right)
\Big]
\label{ss}
\end{eqnarray}
The gauge invariance of the effective action then requires that the
terms on the right-hand side of Eq.\ (\ref{ss}) be invariant under the
gauge transformation $\Omega_{\mu}^{3\,{\rm int}} \to
\Omega_{\mu}^{3\,{\rm int}}+ \partial_{\mu} \Lambda$, which leads to
the condition $b^{\mu \nu \lambda}_{1} = -b^{\nu \mu \lambda}_{1}
$. Consequently, the first two terms in $S_2$ exactly cancel
$S_3$. The property $b^{\mu \nu \lambda}_{1} = -b^{\nu \mu
\lambda}_{1} $ can also be checked by explicit evaluation of the
coefficient $b^{\mu \nu \lambda}_{1}$. However, it is to be noted
that this property and the consequent cancelation of terms in $S_2$
and $S_3$ follows from general gauge invariance requirement and does
not depend on the details of the explicit expressions of the
coefficients $b^{\mu \nu \lambda}_{1}$.

\section {Appendix C}

In this section, we sketch the method of computation of the 
finite temperature polarization tensor $P^{\mu \nu}_3$  which 
is given by
                   \begin{eqnarray}
P^{\mu \nu} _{3} ({\bf p},p_0) &=& \frac{-1}{4 L_x L_y}
{\rm Tr} \left( \int d^2 r_1 \,d^2 r_2  \frac{1}{\beta} \sum_{i \omega_n} 
\frac{1}{2} \left[ \sigma_3 e^{i {\bf p}\cdot {\bf r_1} },
\left(\frac{\partial G_0^{-1}}{\partial \alpha_{\mu}}\right)
_{{\bf r}_1} \right]_+ 
G_0 ({\bf r_1},{\bf r_2},i\omega_n) \right. \nonumber\\
&&\left. \times 
\frac{1}{2} \left [ \sigma_3 e^{-i {\bf p}\cdot {\bf r_2}}, 
\left(\frac{\partial G_0^{-1}}{\partial \alpha_{\nu}}\right)
_{{\bf r}_2} \right]_+ 
G_0 ({\bf r_2},{\bf r_1},i\omega_n + ip_n) \right ).
                     \end{eqnarray}      
The evaluation of $P^{\mu \nu}_3$ is tedious but straightforward. Here
we shall only show the derivation of $P^{00}_3$ explicitly. After
evaluating the matrix trace, we can write $P^{00}_3$ as
                     \begin{eqnarray}
P^{00}_3 &=&  \frac{-1}{4}
\int d^2 r_1 \,d^2 r_2  e^{i {\bf p}\cdot ({\bf r_1}-{\bf r_2})}
\frac{1}{\beta} \sum_{i \omega_n} 
\left( G_0 ^+({\bf r_1},{\bf r_2},i\omega_n) 
G_0 ^+({\bf r_1},{\bf r_2},i\omega_n +ip_n) 
\right. \nonumber\\
&& \left.  
+ G_0 ^-({\bf r_1},{\bf r_2},i\omega_n) 
G_0 ^-({\bf r_1},{\bf r_2},i\omega_n +ip_n) \right).
\label{Poltens}
\end{eqnarray}
The Green Functions $G_0 ^{\pm}$ is easily obtained as solution of
Eq.\ (\ref{geq}) exactly as in the zero temperature case in terms of
the single-particle eigenfunctions and energies (\ref{eigenfunction})
by replacing $\omega \to i\omega_n$:
                 \begin{eqnarray}
G_0 ^{\pm} &=& \frac{|n\rangle \langle n|}
{i \omega_n - \epsilon_n^{\pm}}.
\label{Greenfunction}
                  \end {eqnarray}
Using Eq.\ (\ref{Greenfunction}), it is straightforward to perform the
frequency sums in Eq.\ (\ref{Poltens}). After carrying out the
frequency sum, we perform an analytic continuation to real frequency (
$ ip_n \to p_o + i\eta$ ). It is crucial that the analytic
continuation is performed after the frequency sums are carried out. As
a result, one gets
                    \begin{eqnarray}
P^{00}_3 &=& -\frac{1}{4}\sum_{n=0}^{\infty}\sum_{n'=0}^{\infty}
\sum_{\alpha = \pm} \frac{f_n^{\alpha} - f_{n'}^{\alpha}}
{ p_0 -\omega_c (n-n') + i\eta } 
\langle n | e^{i{\bf p}\cdot {\bf r_1} }|n'\rangle 
\langle n' | e^{-i{\bf p}\cdot {\bf r_2} }|n\rangle 
\,\,\,\,\, {\rm if} \,\, p_0 \ne 0, \nonumber\\
&=& \frac{1}{4}\sum_{n=0}^{\infty}\sum_{n'=0}^{\infty}
\sum_{\alpha = \pm} \delta_{nn'}
\left(-\frac{\partial f(z)}{\partial z}\right)_{z =
\epsilon_n^{\alpha}} \langle n | e^{i{\bf p}\cdot {\bf r_1} }|n'\rangle 
\langle n' | e^{-i{\bf p}\cdot {\bf r_2} }|n\rangle 
\,\,\,\,\, {\rm if} \,\, p_0 =  0,
\label{p00}
                      \end{eqnarray}
where $f_n^{\alpha}$ is shorthand notation for
$f(\epsilon_n^{\alpha})$ and $\epsilon_n^{\alpha}$ are the energy
eigenvalues given by $\epsilon_n^{\alpha} = \epsilon_n +\alpha
\gamma_0$. 

Using the expression for the Landau wavefunctions $\langle n|{\bf
r}\rangle$ \cite{Lopez,Sakhi}, we get for $n \ge n'$
                       \begin{eqnarray} 
\langle n | e^{i{\bf p}\cdot {\bf r_1} }|n'\rangle &=& 
\sqrt{\frac{n'!}{n!}} e^{-\frac{|{\bf p}|^2 l_B^2}{4}}
( p_x + ip_y )^{n-n'} L_{n'}^{n-n'}
(\frac{|{\bf p}|^2 l_B^2}{2}),
\label{expectationvalue}
                      \end{eqnarray}
where $ L_{n'}^{n-n'}$ are the associated Laugurre polynomials.

Using Eq.\ (\ref{expectationvalue}), it is now a matter of
straightforward algebraic manipulations to obtain an expression for
$P_3^{00}$. It is to be noted that in Eq.\ (\ref{p00}) one needs to
separate the terms in the sum with $n \ge n'$ from the terms with $n
\le n'$ and interchange $n$ and $n'$ in the latter. This gives the final
expression for $P_3^{00}$ as
                     \begin{eqnarray}
P_3^{00} &=& |{\bf p}|^2 \Pi_0({\bf p},p_0), \nonumber\\
\Pi_0({\bf p},p_0) &=& \frac{-l_B^2 }{4} 
\sum_{n = 0}^{\infty} \sum_{n' = 0}^{n}
\sum_{\alpha = \pm} C_{nn'}^{\alpha}(p_0)
\frac{n'!}{ n!} e^{\frac{-|{\bf p}|^2 l_B^2}{2}}
\left(\frac{|{\bf p}|^2 l_B^2}{2}\right)^{n-m-1} 
L_{n'}^{n-n'}\left(\frac{|{\bf p}|^2 l_B^2}{2}\right),
                    \end{eqnarray}
where $C_{nn'}^{\alpha}(p_0)$ is defined in Eq.\ (\ref{freqsum}).
The other components of $P_3^{\mu \nu}$ can be computed in an
identical manner and in this way we obtain Eqs.\ (\ref{pmunu}) and 
(\ref{pifunction}). 
 
\section{Appendix D}

In this appendix, we list the notations used in this paper for the
sake of clarity. \\ $\alpha_{\mu}$ :- $ ( \omega,\phi_x,\phi_y )$. \\
$\phi_x,\phi_y$ :- Dimensionless boundary value parameters. \\ ${\bf
H}_0$ :- Constant external magnetic field along z axis. \\ ${\bf
H}_{\rm eff}$ :- Effective magnetic field experienced by the composite
Fermions. \\ ${\bf A}$ :- Vector potential corresponding to ${\bf
H}_0$. \\ $ A_{\mu}^{\rm cs}$ :- Chern-Simons fields. \\ $A_{\mu}^{\rm
em}$ :- External electromagnetic field. \\ $U({\bf r},t)$ :- $2 \times
2$ SU(2) rotation matrices. \\ $\gamma_0$ :- Typical Coulomb energy of
the system $ \sim e^2/l_B$. \\ $\nu$ :- Filling
fraction. \\ $2s$ :- Number of flux quanta attached to electrons. \\ I
:- $2 \times 2$ unit matrix. \\ ${\bbox \sigma}$:- Pauli matrices. \\
${\bf J}$ :- Spin of the skyrmion. \\ ${\bf J}^{\rm Berry}$ :-
Contribution to the spin of the skyrmion from the Berry term. \\
${\bf J}^{\rm Hopf}$ :- Contribution to the spin of the skyrmion from
the bulk Hopf term. \\${\bf J}^{\rm edge}$ :- Contribution  to the
spin of the skyrmion from the edge. \\${\bf j}_0$ :- Spin density of the
skyrmion. \\ $\rho$ :- Charge density of the skyrmion. \\ $\rho_0$ :-
Uniform ground-state particle density. \\
$Q_{\mu}^{\rm int}$ :- Internal gauge field = $ -i U^{-1}
(\partial_{\mu} U)$ = $\frac{1}{2} {\bbox \sigma} \cdot {\bbox \Omega}
_{\mu}^{\rm int} $. \\ $Q_{\mu}^{\theta}$:- Gauge fields due to
rotation by an arbitrary solid angle ${\bbox \theta}$ = $\frac{1}{2}
{\bbox \sigma} \cdot \partial_{\mu} {\bbox \theta}$.\\ $Q_{\mu}^{\rm
total}$ :- Total gauge field = $Q_{\mu}^{\rm int} + U Q_{\mu}^{\bbox
\theta}U^{-1}$.\\ $Q_{\mu}^{\rm eff}$ :- Effective gauge field for
fractional filling = $Q_{\mu}^{\rm int} + A_{\mu}^{\rm cs} I$. \\
$Q_{\rm em}$ :- Electromagnetic Charge of the skyrmion. \\ $Q_{\rm
top}$ :- Topological Charge of the skyrmion. \\ ${\bf d}$ :- Unit
vector field. \\ $f_{\mu \nu}^{\rm int}$ :- Field tensor for the
internal gauge field = 0. \\ $B_{\mu}$ :- Auxiliary gauge field. \\
$F_{\mu \nu}$ :- Field tensor corresponding to the ${\bf B}$ field = $
\partial_{\mu} B_{\nu} - \partial_{\nu} B_{\mu}$ = ${\bf d} \cdot
(\partial_{\mu}{\bf d} \times \partial_{\nu}{\bf d})$. \\ $F_{\mu
\nu}^{\rm cs}$ :- Field tensor corresponding to the ${\bf A}^{\rm cs}$
field. \\ $F_{\mu \nu}^{\rm em}$ :- Field tensor corresponding to the
${\bf A}^{\rm em}$ field. \\ $G_0$ :- Unperturbed Green function. \\
$P^{\mu \nu} _{ij} $ :- The Polarization Bubble. \\ $T^{\mu \nu
\lambda}_{ijk}$ :- The Triangle Diagram. \\ $b_{ij}^{\mu \nu \lambda}$
:- Coefficient of the second order term in the effective action
corresponding to i and j spin indices and $\mu$ $\nu$ and $\lambda$
space-time indices. \\ $b_{j}^{\mu \nu \lambda}$ :- $b_{ij}^{\mu \nu
\lambda}$ with i = j. \\ $c_{ijk}^{\mu \nu \lambda}$ :- Coefficient of
the third order term in the effective action corresponding to i j and
k spin indices and $\mu$ $\nu$ and $\lambda$ space-time indices. \\
$N$ :- prefactor of the Hopf term. \\ $N_0$ :- Number of particles in
the ground state. \\ $\Lambda$ :- Arbitrary
space-time dependent angle corresponding rotation about the ${\bf d}$
axis. \\ ${\bbox \theta}$ :- Arbitrary space-time dependent solid
angle. \\ $\eta$ :- Boson fields for the edge action. \\ $L_x,L_y$ :-
Dimensions of the sample. \\

\begin{figure}
\centerline{\psfig{file=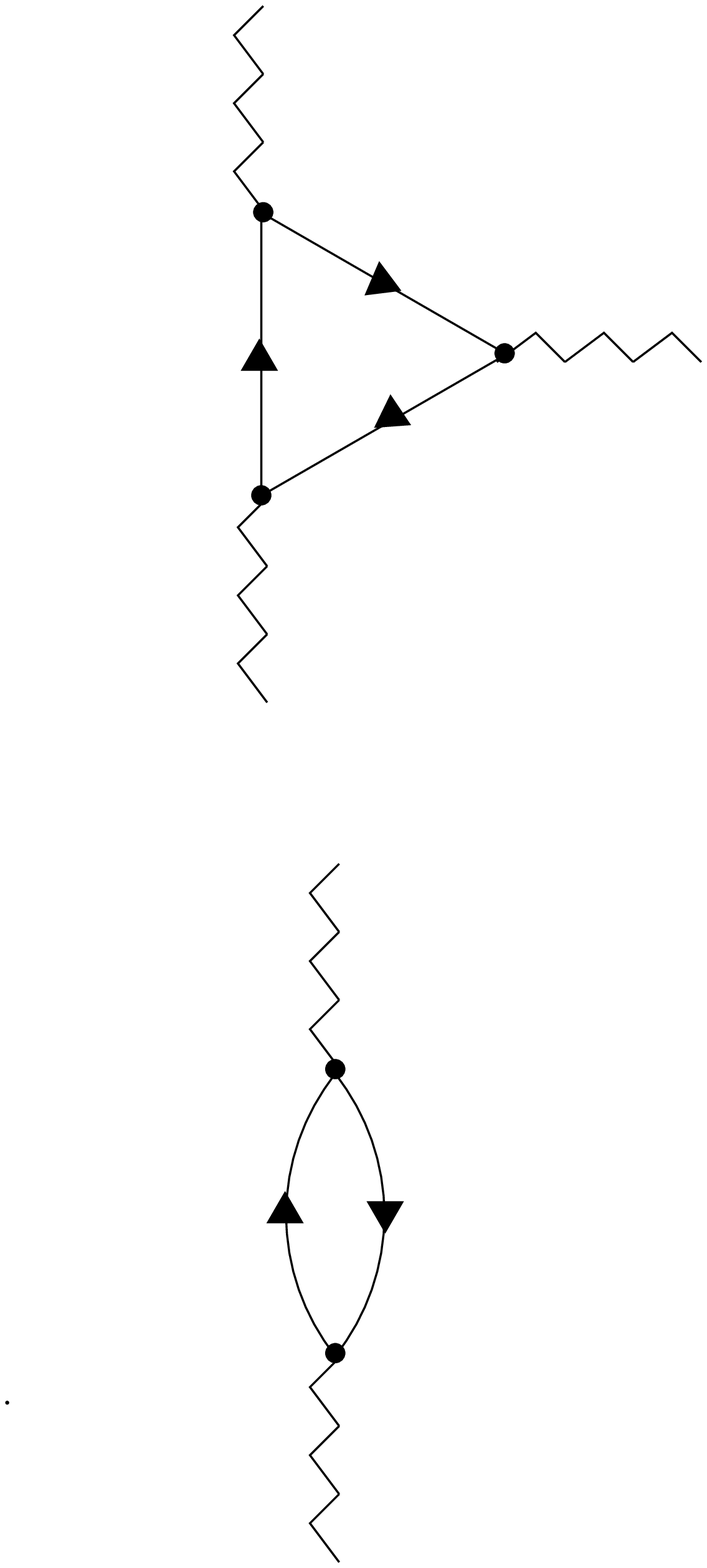,width=\linewidth,angle=-90}}
\caption{Polarization Bubble and Triangular Diagram. The wavy lines 
represent the gauge fields $Q_{\mu}$, the solid lines represent the
Green function $G_0$ and the dots represent the vertices $\frac{\partial
G_0^{-1}}{\partial \alpha_{\mu}}$.}
\label{Graph1}
\end{figure} 
  
\begin{figure}
\centerline{\psfig{file=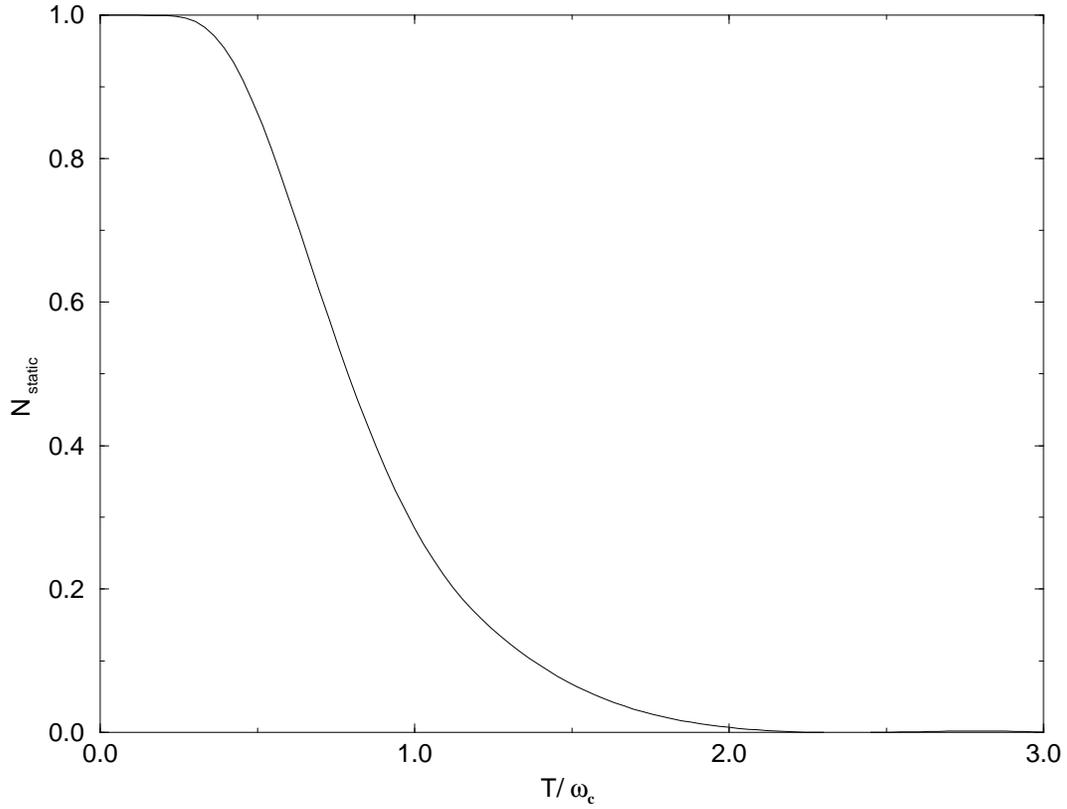,width=\linewidth,angle=-90}}
\caption{Temperature dependence of $N_{\rm static}$.}
\label{Graph2}
\end{figure}           

\end {document}